\documentstyle[aps,prd]{revtex}\tighten

\begin{document}

\def \D {\mbox{D}}
\def \d {\mbox{d}}
\def \t {\tilde}
\def\div {\mbox{div}\,}
\def\rd {\displaystyle{\cdot}}
\def\c {\mbox{curl}\,}
\def\ep {\varepsilon}
\def \ub {u_{\!_B}}
\def \vb {v_{\!_B}}
\def \eb {e_{\!_B}}
\def \st {\sigma_{\!_T}}
\def \ne {n_{\!_E}}
\def \ts {\textstyle}
\def\be {\begin{equation}}
\def\ee {\end{equation}}
\def\bea {\begin{eqnarray}}
\def\eea {\end{eqnarray}}
\def\la {\langle}
\def\ra {\rangle}
\def\p  {\partial}
\def\bi {\bibitem}
\def\case#1/#2{\textstyle\frac{#1}{#2} }

\title{Cosmic microwave background
anisotropies:\\ Nonlinear dynamics
}

\author{
Roy Maartens\dag\thanks{roy.maartens@port.ac.uk}\,,
Tim Gebbie\ddag\,
and George F.R. Ellis\ddag\
}
\address{~}
\address{\dag\ School of Computer Science and Mathematics, Portsmouth
University, Portsmouth~PO1~2EG, England}

\address{\ddag\ Department of Mathematics and Applied Mathematics,
University of Cape Town, Cape~Town~7701, South Africa}
\address{~}

\maketitle

\begin{abstract}

We develop a new approach to local nonlinear effects in cosmic
microwave background anisotropies, and discuss the qualitative
features of these effects.
New couplings of the baryonic velocity
to radiation multipoles are found, arising from nonlinear Thomson
scattering effects.
We also find a new nonlinear shear effect on small angular scales.
The full set of evolution and constraint equations is
derived, including the nonlinear generalizations of the
radiation multipole hierarchy, and of the dynamics of multi-fluids.
These equations govern radiation anisotropies in any
inhomogeneous spacetime, but their main application is to
second-order effects in a
universe that is close to the Friedmann models.
Qualitative analysis is given here, and quantitative calculations
are taken up in further papers.

\end{abstract}

~\\
{\small to appear Phys. Rev. D {\bf 59} (1999)}


\section{Introduction}

Recent and upcoming advances in observations
of the cosmic microwave background (CMB) radiation are
fuelling the construction of increasingly sophisticated
and detailed models to predict the anisotropy on small angular
scales. Such models require highly specific input in order
to produce numerical results, and they involve intricate
problems of computation.
As a complement to such specific predictive models,
it is also useful to pursue a more qualitative and analytical
investigation of CMB anisotropies. A general qualitative analysis
does not rely on detailed assumptions about the origin of primordial
fluctuations, the density parameters of the background,
reionization and structure formation history,
etc. Instead, the aim is to better understand
the underlying physical and geometric factors in the dynamics
of radiation anisotropies, and hopefully to uncover new results
and insights.
In this paper, we follow such an approach,
and develop a new analysis of local nonlinear effects in CMB
anisotropies. We are able to give a physically transparent
qualitative analysis of how inhomogeneities and relative motions
produce nonlinear effects in CMB anisotropies.
We derive the nonlinear generalization of Thomson
scattering, and we find a new nonlinear shear effect on
small scales.

We use a 1+3 covariant approach (i.e., a ``covariant Lagrangian"
approach) to CMB anisotropies, based on choice of
a physically determined 4-velocity vector field $u^a$.
This allows us to derive the exact nonlinear equations for physical
quantities as measured by observers moving with that 4-velocity.
Then the nonlinear equations
provide a covariant basis for investigating second-order effects, as
well as for linearizing about a Friedmann-Lemaitre-Robertson-Walker
(FLRW) background. The basic theoretical ingredients are:
(a) the covariant Lagrangian dynamics of Ehlers and
Ellis \cite{ehl,ell}, and the perturbation theory of
Hawking \cite{haw}
and Ellis and Bruni \cite{eb} which is derived from it;
(b) the 1+3 covariant
kinetic theory formalism of Ellis, Treciokas and
Matravers \cite{ETMa,ETMb} (which builds on work by Ehlers, Geren and
Sachs \cite{EGS}, Treciokas and Ellis \cite{TE} and Thorne \cite{T2});
and (c) the 1+3 covariant analysis of temperature anisotropies due
to Maartens, Ellis and Stoeger \cite{mes}.

The well-developed study of CMB anisotropies is
based on the pioneering
results in CMB physics (Sachs and Wolfe \cite{SW},
Rees and Sciama \cite{rs}, Peebles and Yu \cite{PY},
Sunyaev and Zeldovich \cite{sz2},
Grishchuk and Zeldovich \cite{gz},
and others), and on the development of gauge-invariant
perturbation theory,
particularly by Bardeen \cite{bar} and Kodama and Sasaki \cite{ks}
(building of the work of Lifshitz \cite{lif}).
There are comprehensive and detailed
models -- see e.g. Hu and Sugiyama \cite{HS95a,HS95b,HW}, Ma and
Bertschinger \cite{MB}, Seljak et al. \cite{sz,zs,zsb,hswz},
Durrer and Kahniashvili \cite{dk}.
These provide the basis for sophisticated predictions and
comparisons with the observations of recent, current and
future satellite and ground-based experiments.
The hope is that this inter-play between theory and observation
(including the large-scale galactic distribution and
other observations),
in the context of inflationary cosmology, will produce
accurate values for the various parameters that characterize
the standard models, thus allowing theorists to discriminate
between competing models (see for example \cite{gs,hrlg}).

While these papers have provided a near-exhaustive
treatment of many of the issues involved in CMB physics,
there are a number of reasons for pursuing a
complementary 1+3 covariant approach, as developed in
\cite{mes,sme,mes3,mes4,SAG,dunsby,cl,cl2,GE,c}.

Firstly, the covariant approach by its very nature incorporates
nonlinear effects.
This approach starts from the inhomogeneous and anisotropic universe,
without a priori restrictions on the degree of inhomogeneity
and anisotropy,
and then applies the linearization limit when required. The
1+3 covariant equations governing CMB anisotropies are thus
applicable in fully nonlinear generality. These equations
can then be specialized in various ways in addition to a standard
FLRW-linearization. Second-order effects
in an almost-FLRW universe probably
form the most important possibility, given the increasing accuracy
and refinement of observations. The study of CMB anisotropies
in homogeneous Bianchi universes with large anisotropy is another
possibility that flows directly from the general nonlinear equations.
Such applications will be the subject of future papers in
the programme. The current paper is concerned with setting up the
general dynamical equations and identifying the qualitative
nature of nonlinear effects. (The general algebraic equations
are derived in \cite{GE}.)

Secondly, the 1+3 covariant approach is based entirely on quantities
with a direct and transparent physical and geometric interpretation,
and the fundamental quantities describing anisotropy and inhomogeneity
are all automatically gauge-invariant when a suitable
covariant choice of
fundamental 4-velocity has been made. As a consequence the approach
leads to results with unambiguous physical meaning (provided the
fundamental 4-velocity field is chosen in a physically unique and
appropriate way; we discuss the various options below).

This approach has been developed in the
context of density perturbations
\cite{eb,ehb,ebh,ls,dun2,bde,bed,dbe,bggmv,ber,mt,tb,mt2,vee,maa2,bm,mtm}
and gravitational wave perturbations \cite{haw,dbe2,he,mes2,mb2,maa3}.
(See also \cite{ed} for a recent review.) In
relation to CMB anisotropies, the covariant Lagrangian
approach was initiated\footnote{
A 1+3 covariant approach to CMB anisotropy was independently
outlined by Bonanno and Romano \cite{br} in general terms,
using a flux-limited diffusion theory, but
the detailed implications of small CMB anisotropy were not pursued.
}
by Stoeger, Maartens and Ellis \cite{sme}, who proved the following
result:
{\em if all comoving observers in an expanding universe region
measure the anisotropy of the CMB after last scattering to be small,
then the universe is almost FLRW in that region.}\footnote{
Note the importance of expansion: a {\em static} isotropic cosmology with
arbitrarily large inhomogeneity can be
constructed in which all observers see isotropic CMB \cite{emn}.
}
No a priori assumptions are made on the spacetime geometry,
or on the source and nature of CMB anisotropies, so that
this result provides a general theoretical underpinning for CMB
analysis in perturbed FLRW universes. It effectively
constitutes a proof of the stability of the corresponding
exact-isotropy result of Ehlers, Geren and Sachs \cite{EGS}.
The weak Copernican principle implicit in the assumption that
all fundamental observers see small anisotropy is in principle
partially testable via the Sunyaev-Zeldovich effect
(see \cite{mes4} and references therein).
The qualitative result was extended into a quantitative
set of limits on the anisotropy and inhomogeneity of the
universe imposed by the observed degree of CMB anisotropy,
independently of any assumptions on cosmic dynamics
or perturbations before recombination \cite{mes,mes3,mes4,SAG}.

More recently, this approach to CMB anisotropies in
an almost FLRW universe has been extended by Dunsby \cite{dunsby},
who derived a 1+3 covariant version of the Sachs-Wolfe formula,
and by Challinor and Lasenby \cite{cl,cl2}, who performed a
comprehensive 1+3 covariant
analysis of the imprint of scalar perturbations
on the CMB, confirming the results of other
approaches from this viewpoint and bringing new insights and
clarifications via the covariant approach. In \cite{cl},
they also discuss qualitatively
the imprint of tensor perturbations on the CMB, in the covariant
approach (see \cite{c} for quantitative results).

This paper is closely related to,
and partly dependent upon, all of these previous
1+3 covariant analyses.
It extends and generalizes aspects of these papers,
using and developing the covariant nonlinear
Einstein-Boltzmann-hydrodynamic formalism.
We analyze the nonlinear dynamics of radiation
anisotropies, with the main application being
second-order effects in an almost FLRW universe.
We identify and describe the qualitative features of
such effects. This lays the basis for a generalization of
results on well known second-order effects such as the Rees-Sciama
and Vishniac effects (see e.g. \cite{HS95a}),
and on recent second-order corrections
of the Sachs-Wolfe effect \cite{pc,mm2}.
Developing a quantitative analysis on the basis
of the equations and qualitative analysis given here
is the subject of further work.
Ultimately this
involves the solution
of partial differential equations, which
requires in particular a choice of coordinates, breaking
covariance. However, the 1+3 covariant approach means that
all the equations and variables have a
direct and transparent physical meaning.

In Section II, the covariant Lagrangian
formalism for relativistic cosmology is briefly summarized.
Section III develops
an exact 1+3 covariant treatment of multi-fluids and
their relative velocities, building on \cite{maa2}.
In Section IV, the covariant Lagrangian approach to kinetic theory
is outlined. Section V develops a nonlinear treatment of
Thomson scattering, which identifies new couplings of the baryonic
relative velocity to the radiation multipoles.
We derive the hierarchy of exact covariant multipole equations
which arise from the Boltzmann equation.
This section uses and generalizes a combination of the
results of Ellis et al. \cite{ETMa} on the multipoles of
the Boltzmann equation in general, Maartens et al. \cite{mes}
on a covariant description of temperature fluctuations,
and Challinor and Lasenby \cite{cl2} on Thomson scattering.
The equations constitute a covariant and nonlinear generalization
of previous linearized treatments.
In Section VI, we consider qualitative
implications of the nonlinear equations.
We identify the role of the kinematic
quantities in the nonlinear terms, and comment on the implications
for second-order effects,
which include a new nonlinear shear correction
to CMB anisotropies on small angular scales. We also
give the multipole equations for the case
where the radiation anisotropy
is small, but spacetime anisotropy and inhomogeneity
are unrestricted.\footnote{
This case will apply before decoupling, in order to be
consistent with the almost-FLRW result quoted above.
}

Finally, we give the linearized form of the multipole equations,
regaining the equations of Challinor and Lasenby \cite{cl2}.
This provides a covariant Lagrangian
version of the more usual
metric-based formalism of gauge-invariant
perturbations (see e.g. \cite{W83,HS95b,MB,sz}).
In a further paper \cite{ge2}, the linearized equations derived here
are expanded in covariant scalar modes, and this is used
to determine analytic properties of CMB linear anisotropy formation.

We follow the notation and conventions of \cite{ell,ETMa,mes}, with
the improvements and developments introduced by \cite{maa,mt}. In particular:
the units are such that $c$, $8\pi G$ and $k_{\!_B}$ are equal 1;
the signature is $(-+++)$;
spacetime indices are $a,b,\cdots\,=0,1,2,3$;
the curvature tensors are $R^a{}_{bcd}=-\p_d\Gamma^a{}_{bc}+\cdots$,
$R_{ab}=R^c{}_{acb}$ and $R=R^a{}_a$, and the Ricci identity is
$\nabla_{[a}\nabla_{b]}u_c={1\over2}R_{abcd}u^d$;
$A_\ell$ denotes the
index string $a_1a_2\cdots a_\ell$, and ${e}^{A_\ell}$ denotes
the tensor product $e^{a_1}e^{a_2}\cdots e^{a_\ell}$;
(square) round brackets
enclosing indices denote the (anti-)symmetric part, and angled
brackets denote the projected symmetric and tracefree (PSTF) part
(defined below).
The spatially projected part of the covariant derivative is denoted
by $\D_a$, following \cite{maa}.\footnote{
In \cite{eb,bde,dbe,cl,cl2} it is
denoted $^{(3)}\nabla_a$, while in \cite{mes,dunsby} it is
$\widehat{\nabla}_a$.
}
The approximate equality symbol,
as in $J\approx 0$, indicates
equality up to first (linear) order in an almost-FLRW spacetime.

\section{Covariant Lagrangian formalism in relativistic cosmology}

The Ehlers-Ellis 1+3 formalism \cite{ehl,ell,kt}
is a covariant Lagrangian approach, i.e. every quantity has a
natural interpretation in terms of observers comoving with the
fundamental 4-velocity $u^a$ (where $u^au_a=-1$).
Provided this is defined uniquely in an invariant manner,
all related quantities have a direct physical or geometric meaning,
and may in principle be measured in the instantaneous rest space
of the comoving fundamental observers.
Any coordinate system or tetrad can be used when specific
calculations are made. These features are a crucial part of
the strengths of the formalism and of
the perturbation theory that is derived from it.
We will follow the streamlining and development of the formalism
given by Maartens \cite{maa}, the essence of which is to
make explicit use of irreducible quantities and derivatives, and
to develop the identities which these quantities and derivatives
obey (see also \cite{mt,mes2,mb2,vee,maa2}).

The basic algebraic tensors are: (a)
the projector $h_{ab}=g_{ab}+u_au_b$, where $g_{ab}$ is the spacetime
metric, which projects into the instantaneous rest space
of comoving observers;
and (b) the projected alternating tensor
$\ep_{abc}=\eta_{abcd}u^d$, where $\eta_{abcd}=-\sqrt{|g|}
\delta^0{}_{[a}\delta^1{}_b\delta^2{}_c\delta^3{}_{d]}$ is the
spacetime alternating tensor. Thus
\[
\eta_{abcd} = 2u_{[a}\ep_{b]cd}-2\ep_{ab[c}u_{d]}\,,~
\ep_{abc}\ep^{def}=3!h_{[a}{}^dh_b{}^eh_{c]}{}^f\,.
\]
The projected symmetric tracefree (PSTF)
parts of vectors and rank-2 tensors are
\[
V_{\la a\ra}=h_a{}^bV_b\,,~
S_{\la ab\ra }= \left\{h_{(a}{}^ch_{b)}{}^d-
{\ts{1\over3}}h^{cd}h_{ab}\right\}S_{cd}\,,
\]
with higher rank formulas given in \cite{GE}. The skew part of
a projected rank-2 tensor is spatially dual to the projected
vector $S_a={1\over2}\ep_{abc}S^{[bc]}$, and then any projected
rank-2 tensor has the irreducible covariant decomposition
\[
S_{ab}=\case{1}/{3}Sh_{ab}+\ep_{abc}S^c+S_{\la ab\ra}\,,
\]
where $S=S_{cd}h^{cd}$ is the spatial trace. In the 1+3 covariant
formalism, all quantities are either scalars, projected vectors
or PSTF tensors.
The equations governing these quantities involve a covariant
vector product and its generalization to PSTF rank-2 tensors:
\[
[V,W]_a=\ep_{abc}V^bW^c\,,~[S,Q]_a=\ep_{abc}S^b{}_dQ^{cd}\,.
\]

The covariant derivative $\nabla_a$ defines
1+3 covariant time and spatial derivatives
\[
\dot{J}^{a\cdots}{}{}_{\cdots b}= u^c \nabla_c
J^{a\cdots}{}{}_{\cdots b}\,,~
\D_c J^{a\cdots}{}{}_{\cdots b} =
    h_c{}^d h^a{}_e\cdots h_b{}^f
    \nabla_d J^{e\cdots}{}{}_{\cdots f}\,.
\]
Note that $\D_ch_{ab}=0=\D_d\ep_{abc}$, while
$\dot{h}_{ab}=2u_{(a}\dot{u}_{b)}$ and $\dot{\ep}_{abc}=
3u_{[a}\ep_{bc]d}\dot{u}^d$. The projected derivative $\D_a$
further splits irreducibly into a
1+3 covariant spatial divergence and curl \cite{maa}
\begin{eqnarray*}
&& \div V=\D^aV_a\,,~(\div S)_a=\D^bS_{ab}\,, \\
&& \c V_a=\ep_{abc}\D^bV^c\,,~ \c S_{ab}=\ep_{cd(a}\D^cS_{b)}{}^d\,,
\end{eqnarray*}
and a 1+3 covariant spatial distortion \cite{mes2}
\begin{eqnarray*}
\D_{\la a}V_{b\ra}&=&\D_{(a}V_{b)}-{\ts{1\over3}}
\left(\div V\right)h_{ab}\,,\\
\D_{\la a}S_{bc\ra}
&=&\D_{(a}S_{bc)}-{\ts{2\over5}}h_{(ab}\left(\div S\right)_{c)}\,.
\end{eqnarray*}
Note that div curl is {\em not} in general zero, for vectors
or rank-2 tensors (see \cite{maa,mt,vee,mb2}
for the relevant formulas).
The covariant irreducible decompositions of the derivatives of
scalars, vectors and rank-2 tensors are given in exact (nonlinear)
form by \cite{mes2}
\begin{eqnarray}
\nabla_a \psi &=& -\dot{\psi}u_a+\D_a \psi \,,\label{a1}\\
\nabla_bV_a &=&-u_b\left\{
\dot{V}_{\la a\ra}+
A_cV^cu_a\right\}
+u_a\left\{{\ts{1\over3}}\Theta  V_b+\sigma_{bc}V^c
+[\omega,V]_b\right\} \nonumber\\
&&{}+{\ts{1\over3}}\left(\div V\right)h_{ab}-{\ts{1\over2}}\ep_{abc}\c
V^c+\D_{\la a}{V}_{b\ra}\,, \label{a2}\\
\nabla_cS_{ab} &=& -u_c\left\{\dot{S}_{\la ab\ra}+2u_{(a}
S_{b)d}A^d\right\}+2u_{(a}\left\{{\ts{1\over3}}\Theta S_{b)c}
+S_{b)}{}^d\left(\sigma_{cd}-\ep_{cde}\omega^e\right)\right\}
\nonumber\\
&&{}+{\ts{3\over5}}\left(\div S\right)_{\la a}h_{b\ra c}
-{\ts{2\over3}}\ep_{dc(a}\c S_{b)}{}^d+\D_{\la a}{S}_{bc\ra}
\,. \label{a3}
\end{eqnarray}

The algebraic correction terms in equations
(\ref{a2}) and (\ref{a3}) arise from the
relative motion of comoving observers, as encoded in
the kinematic quantities:
the expansion
$\Theta=\D^au_a$, the 4-acceleration
$A_a\equiv\dot{u}_a=A_{\la a\ra }$, the vorticity\footnote{
The vorticity tensor $\omega_{ab}=\ep_{abc}\omega^c$ is often used,
but we prefer to use the irreducible vector
$\omega_a$. The sign conventions, following \cite{ehl,ell},
are such that in the Newtonian limit, $\vec{\omega}=-{1\over2}
\vec{\nabla}\times \vec{v}$. Note that $\D^b\omega_{ab}=
\c\omega_a$.
}
$\omega_a=-{\ts{1\over2}}\c u_a$, and the shear
$\sigma_{ab}=\D_{\la a}u_{b\ra }$.
Thus, by Eq. (\ref{a2})
\[
\nabla_bu_a=-A_au_b+{\ts{1\over3}}\Theta h_{ab}+\ep_{abc}\omega^c
+\sigma_{ab}\,.
\]

The irreducible parts of the Ricci identities produce commutation
identities for the irreducible derivative operators.
In the simplest case of scalars:
\begin{eqnarray}
\c\D_a\psi&\equiv& \ep_{abc}\D^{[b}\D^{c]}\psi=
-2\dot{\psi}\omega_a \,, \label{ri1}\\
\D_a\dot{\psi}-h_a{}^b\left(\D_b\psi\right)^{\rd} &=&-\dot{\psi}A_a
+{\ts{1\over3}}\Theta\D_a\psi+\sigma_a{}^b\D_b\psi+[\omega,\D\psi]_a
\,.\label{ri2}
\end{eqnarray}
Identity (\ref{ri1})
reflects the crucial relation of vorticity to non-integrability;
non-zero $\omega_a$ implies that there are no constant-time
3-surfaces everywhere orthogonal to $u^a$, since the instantaneous
rest spaces cannot be patched together smoothly.\footnote{
In this case,
which has no Newtonian counterpart, the $\D_a$ operator is not
intrinsic to a 3-surface, but it is still a well-defined spatial
projection of $\nabla_a$ in each instantaneous rest space.
}
Identity (\ref{ri2}) is the key to deriving evolution equations
for spatial gradients, which covariantly characterize inhomogeneity
\cite{eb}.
Further identities are given in \cite{ebh,maa,ve,mes2,mt}.

The kinematic quantities govern the relative motion of
neighboring fundamental world-lines, and describe the universal
expansion and its local anisotropies.
The dynamic quantities describe the sources of the gravitational
field, and directly determine the Ricci curvature locally via
Einstein's field equations.
They are the
(total) energy density $\rho=T_{ab}u^au^b$, isotropic pressure
$p={1\over3}h_{ab}T^{ab}$, energy flux $q_a=-T_{\la a\ra b}u^b$,
and anisotropic stress $\pi_{ab}=T_{\la ab\ra}$, where $T_{ab}$
is the total energy-momentum tensor. The
locally free gravitational field,
i.e. the part of the spacetime
curvature not directly determined locally
by dynamic sources, is given
by the Weyl tensor $C_{abcd}$. This splits irreducibly into the
gravito-electric and gravito-magnetic fields
\[
E_{ab}=C_{acbd}u^cu^d=E_{\la ab\ra }\,,~~
H_{ab}={\ts{1\over2}}\ep_{acd}C^{cd}{}{}_{be}u^e=H_{\la ab\ra} \,,
\]
which provide a covariant Lagrangian description of
tidal forces and gravitational radiation.

An FLRW (background) universe, with its unique preferred 4-velocity
$u^a$, is covariantly characterized as follows:\\
dynamics: $\D_a\rho=0=\D_a p$,
$q_a=0$, $\pi_{ab}=0$;\\
kinematics: $\D_a\Theta=0$, $A_a=0=\omega_a$,
$\sigma_{ab}=0$;\\
gravito-electric/magnetic field: $E_{ab}=0=H_{ab}$.

The Hubble rate
is $H={1\over3}\Theta=\dot{a}/a$, where $a(t)$ is the scale
factor and $t$ is cosmic proper time. In spatially homogeneous but
anisotropic universes (Bianchi and Kantowski-Sachs models),
the quantities $q_a$, $\pi_{ab}$, $\sigma_{ab}$, $E_{ab}$
and $H_{ab}$ in the preceding list may be non-zero.

The Ricci identity for $u^a$ and the Bianchi identities
$\nabla^d C_{abcd} = \nabla_{[a}(-R_{b]c} + {1\over6}Rg_{b]c})$
produce the fundamental evolution and constraint equations
governing the above covariant quantities \cite{ehl,ell}.
Einstein's equations are incorporated\footnote{ Note that
one constraint Einstein equation is not explicitly
contained in this set -- see \cite{ell,mac}.
}
via the algebraic replacement
of the Ricci tensor $R_{ab}$
by $T_{ab}-{1\over2}T_c{}^cg_{ab}$. These
equations, in exact (nonlinear) form and for a general
source of the gravitational field, are \cite{mes2}:\\

\noindent{\sf Evolution:}
\begin{eqnarray}
 \dot{\rho} +(\rho+p)\Theta+\div q&=& -2A^a q_a
 -\sigma^{ab}\pi_{ab} \,, \label{e1}\\
\dot{\Theta} +{\ts{1\over3}}\Theta^2
+{\ts{1\over2}}(\rho+3p)-\div A &=&
 -\sigma_{ab}\sigma^{ab}
+2\omega_a\omega^a+A_aA^a \,,
\label{e2}\\
 \dot{q}_{\la a\ra }
+{\ts{4\over3}}\Theta q_a+(\rho+p)A_a +\D_a p
+(\div\pi)_{a} &=&
-\sigma_{ab}q^b+[\omega,q]_a
-A^b\pi_{ab} \,,
\label{e3} \\
 \dot{\omega}_{\la a\ra } +{\ts{2\over3}}\Theta\omega_a
+{\ts{1\over2}}\c A_a &=&\sigma_{ab}\omega^b \,,\label{e4}\\
 \dot{\sigma}_{\la ab\ra } +{\ts{2\over3}}\Theta\sigma_{ab}
+E_{ab}-{\ts{1\over2}}\pi_{ab} -\D_{\la a}A_{b\ra } &=&
-\sigma_{c\la a}\sigma_{b\ra }{}^c-
\omega_{\la a}\omega_{b\ra }
+A_{\la a}A_{b\ra }\,,
\label{e5}\\
 \dot{E}_{\la ab\ra } +\Theta E_{ab}
-\c H_{ab}
+{\ts{1\over2}}(\rho+p)\sigma_{ab}
&+&{\ts{1\over2}}
\dot{\pi}_{\la ab\ra }
+{\ts{1\over2}}\D_{\la a}q_{b\ra }+{\ts{1\over6}}
\Theta\pi_{ab} \nonumber\\
&=&{}-A_{\la a}q_{b\ra }
 +2A^c\ep_{cd(a}H_{b)}{}^d
+3\sigma_{c\la a}E_{b\ra }{}^c \nonumber\\
&&{}-\omega^c \ep_{cd(a}E_{b)}{}^d
-{\ts{1\over2}}\sigma^c{}_{\la a}\pi_{b\ra c}
-{\ts{1\over2}}\omega^c\ep_{cd(a}\pi_{b)}{}^d \,,
\label{e6}\\
\dot{H}_{\la ab\ra } +\Theta H_{ab}
+\c E_{ab}
-{\ts{1\over2}}\c\pi_{ab}&=&
3\sigma_{c\la a}H_{b\ra }{}^c
-\omega^c \ep_{cd(a}H_{b)}{}^d \nonumber\\
&&{}-2A^c\ep_{cd(a}E_{b)}{}^d
-{\ts{3\over2}}\omega_{\la a}q_{b\ra }+
{\ts{1\over2}}\sigma^c{}_{(a}\ep_{b)cd}q^d \,.
\label{e7}
\end{eqnarray}

\noindent{\sf Constraint:}
\begin{eqnarray}
\div\omega &=&A^a\omega_a \,,\label{c1}\\
(\div\sigma)_{a}-\c\omega_a
-{\ts{2\over3}}\D_a\Theta
+q_a &=&-
2[\omega,A]_a  \,,\label{c2}\\
  \c\sigma_{ab}+\D_{\la a}\omega_{b\ra }
 -H_{ab}&=& -2A_{\la a}
\omega_{b\ra } \,,\label{c3}\\
 (\div E)_{a}
+{\ts{1\over2}}(\div\pi)_{a}
 -{\ts{1\over3}}\D_a\rho
+{\ts{1\over3}}\Theta q_a
&=&[\sigma,H]_a
-3H_{ab}\omega^b
+{\ts{1\over2}}\sigma_{ab}q^b-{\ts{3\over2}}
[\omega,q]_a  \,,\label{c4}\\
 (\div H)_{a}
+{\ts{1\over2}}\c q_a
 -(\rho+p)\omega_a &=&
-[\sigma,E]_a-{\ts{1\over2}}[\sigma,\pi]_a
+3E_{ab}\omega^b -{\ts{1\over2}}\pi_{ab}
\omega^b  \,.\label{c5}
\end{eqnarray}

If the universe is close to an FLRW model, then
quantities that vanish in the FLRW limit are
$O(\epsilon)$, where $\epsilon$ is a
dimensionless smallness parameter, and the quantities are suitably
normalized (e.g. $\sqrt{\sigma_{ab}\sigma^{ab}}/H<\epsilon$, etc.).
The above
equations are covariantly and gauge-invariantly
linearized \cite{eb} by dropping all terms $O(\epsilon^2)$, and
by replacing scalar coefficients of $O(\epsilon)$ terms
by their background values.
This linearization reduces all the right hand sides
of the evolution and constraint equations to zero.

\section{1+3 Covariant nonlinear analysis of multi-fluids}

The formalism described above applies
for {\em any} covariant choice of $u^a$.
If the physics picks out only one $u^a$,
then that becomes the natural and obvious 4-velocity to use.
In a complex
multi-fluid situation, however, there are various possible
choices. The different particle species in cosmology
will each have distinct 4-velocities; we could choose any of these as
the fundamental frame, and
other choices such as the centre of mass frame are also possible.
This allows a variety of covariant choices of 4-velocities,
each leading to a slightly different 1+3 covariant description.
One
can regard a choice between these different possibilities as a partial
gauge-fixing (but determined in a covariant and physical way).
Any differences between such 4-velocities
will be $O(\epsilon)$ in the almost-FLRW case and will
disappear in the FLRW limit,\footnote{
A similar situation occurs in relativistic thermodynamics,
where suitable 4-velocities are close to the equilibrium 4-velocity,
and hence to each other \cite{Is}.
}
as is required in a consistent 1+3 covariant
and gauge-invariant linearization about an
FLRW model (see \cite{eb,cl} for further discussion).

In addition to the issue of linearization,
one can also ask more generally
what the impact of a change of fundamental frame is on the
kinematic, dynamic and gravito-electric/magnetic quantities.
If an initial choice $u^a$ is replaced by a new choice $\t{u}^a$, then
\be
\t{u}^a=\gamma(u^a+v^a)~\mbox{ where }~v_au^a=0\,,~
\gamma=(1-v_av^a)^{-1/2} \,,
\label{r2}\ee
where $v^a$ is the (covariant) velocity of the new frame relative to
the original frame.
The exact transformations of all relevant quantities are
given in the appendix, and are taken from \cite{maa2}.
To linear order, the transformations take the form:
\begin{eqnarray*}
&&\t{\Theta}\approx\Theta+\div v\,,~~\t{A}_a\approx A_a+
\dot{v}_a+Hv_a\,, \\
&&\t{\omega}_a\approx\omega_a-{\ts{1\over2}}\c v_a\,,~~
\t{\sigma}_{ab}\approx\sigma_{ab}+\D_{\la a}v_{b\ra}\,,\\
&&\t{\rho}\approx\rho\,,~~\t{p}\approx p\,,~~\t{q}_a\approx
q_a-(\rho+p)v_a\,,~~\t{\pi}_{ab}\approx\pi_{ab}\,,\\
&&\t{E}_{ab}\approx E_{ab}\,,~~\t{H}_{ab}\approx H_{ab}\,.
\end{eqnarray*}

Suppose now that a choice of fundamental frame has been made.
(For the purposes of this paper, we will not need to
specify such a choice.)
Then we need to consider the velocities of each species
which source the gravitational field, relative to the
fundamental frame.
If the 4-velocities are close, i.e. if the frames are in
non-relativistic relative motion, then
$O(v^2)$ terms may be dropped from the equations,
except if we include nonlinear kinematic, dynamic and
gravito-electric/magnetic effects, in which case, for consistency, we
must retain $O(\epsilon^0v^2)$
terms such as $\rho v^2$, which are of the same order of magnitude
in general as $O(\epsilon^2)$ terms. (See \cite{pl}.)
If the universe is close to FLRW, then $O(\epsilon^0v^2)$
terms may be neglected, together with $O(\epsilon v)$
and $O(\epsilon^2)$ terms.

In summary, there are two different linearizations:

(a) linearizing
in relative velocities (i.e. assuming all species have nonrelativistic
bulk motion relative to the fundamental frame), without
linearizing in the kinematic, dynamic and gravito-electric/magnetic
quantities that covariantly characterize the spacetime;

(b) FLRW-linearization, which implies the special case of (a)
obtained by also linearizing in the kinematic, dynamic and
gravito-electric/magnetic quantities.

Clearly (a) is more general,
and we can take it to be the physically relevant nonlinear regime,
i.e. the case where only
nonrelativistic average velocities\footnote{
Of course, this implies no restrictions on the
velocities of individual particles within any species.
}
are considered, but no other assumptions are made on the
physical or geometric quantities.
In case (a), no restrictions are imposed on non-velocity terms, and
we neglect only terms $O(\epsilon v^2,v^3)$. In
case (b), we neglect terms $O(\epsilon^2,\epsilon v,v^2)$.
Covariant
second-order effects against an FLRW background are included within
(a), when we neglect terms $O(\epsilon^3)$. (Note
that gauge-invariance is a far more subtle problem at second
order than at first order: see Bruni et al. \cite{bmms}.)

The dynamic quantities in the evolution and constraint
equations (\ref{e1})--(\ref{c5}) are the total quantities, with
contributions from all dynamically significant particle species.
Thus
\begin{eqnarray}
T^{ab} &=& \sum_I T_{\!_I}^{ab} = \rho u^au^b+ph^{ab}+2q^{(a}u^{b)}
+\pi^{ab} \,, \label{t1}\\
T_{\!_I}^{ab}&=& \rho_{\!_I}u_{\!_I}^au_{\!_I}^a+p_{\!_I}h_{\!_I}^{ab}
+2q_{\!_I}^{(a}u_{\!_I}^{b)}+\pi_{\!_I}^{ab}\,, \label{t2}
\end{eqnarray}
where $I$ labels the species.
We
include radiation photons ($I=R$),
baryonic matter ($I=B$) modelled as a perfect fluid,
cold dark matter ($I=C$) modelled as dust
over the era of interest for CMB anisotropies, neutrinos
($I=N$) (assumed to be massless),
and a cosmological constant ($I=V$).\footnote{
A more general treatment, incorporating all the sources which
are currently believed to be potentially significant, would
also include
a dynamic scalar field that survives
after inflation (``quintessence"), and hot dark matter in the form
of massive neutrinos (see \cite{gs} for a survey
with further references). Our main aim is not a detailed
and comprehensive model with numerical predictions, but a
qualitative discussion focusing on the underlying dynamic and
geometric effects at nonlinear and linear level
that are brought out clearly by a 1+3 covariant approach.
In principle our approach
is readily generalized to include other sources of the
gravitational field.
}
Note that the dynamic quantities $\rho_{\!_I},\cdots$ in equation
(\ref{t2}) are as measured in the $I$-frame, whose 4-velocity
is given by
\be
u_{\!_I}^a=\gamma_{\!_I}\left(u^a+v_{\!_I}^a\right)\,,~v_{\!_I}^au_a=0
\,.\label{t3}
\ee
Thus we have
\begin{eqnarray}
&& p_{\!_C}=0=q_{\!_C}^a=\pi_{\!_C}^{ab}\,,~~
q_{\!_B}^a=0=\pi_{\!_B}^{ab}\,, \label{t3a}\\
&&p_{\!_R}=\case{1}/{3}\rho_{\!_R}\,,~~p_{\!_N}=\case{1}/{3}
\rho_{\!_N}\,, \label{t3b}
\end{eqnarray}
where we have chosen the unique 4-velocity in the cold dark matter and
baryonic cases which follows from modelling these fluids as perfect.
The cosmological constant is characterized by
\[
p_{\!_V}=-\rho_{\!_V}=-\Lambda\,,~q_{\!_V}^a=0=\pi_{\!_V}^{ab}\,,~~
v_{\!_V}^a=0\,.
\]

The conservation equations for the species are best given
in the overall $u^a$-frame, in terms of the velocities $v_{\!_I}^a$
of species $I$ relative to this frame.
Furthermore, the evolution and constraint
equations of Section II are all given in terms of the
$u^a$-frame.
Thus we need the expressions for
the partial dynamic quantities as measured in the overall frame.
The velocity formula inverse to equation (\ref{t3}) is
\begin{equation}
u^a=\gamma_{\!_I}\left(u_{\!_I}^a+v_{\!_I}^{*a}\right)\,,~~
v_{\!_I}^{*a}=-\gamma_{\!_I}\left(v_{\!_I}^a+v_{\!_I}^2u^a\right)\,,
\label{t3c}\end{equation}
where $v_{\!_I}^{*a}u_{\!_Ia}=0$, and $v_{\!_I}^{*a}v_{\!_Ia}^*=
v_{\!_I}^av_{\!_Ia}$.
Using this relation together with the general transformation
equations (\ref{ab7})--(\ref{ab10}), or directly from the
above equations, we find the following exact (nonlinear)
equations for the dynamic quantities of species $I$ as
measured in the overall $u^a$-frame:
\begin{eqnarray}
\rho_{\!_I}^* &=& \rho_{\!_I}
+ \left\{\gamma_{\!_I}^2v_{\!_I}^2\left(\rho_{\!_I}+p_{\!_I}\right)
+2\gamma_{\!_I}q_{\!_I}^a
v_{\!_Ia}+\pi_{\!_I}^{ab}v_{\!_Ia}v_{\!_Ib}\right\} \,,\label{t4}\\
p_{\!_I}^* &=&  p_{\!_I}
+{\textstyle{1\over3}}
\left\{\gamma_{\!_I}^2v_{\!_I}^2\left(\rho_{\!_I}
+p_{\!_I}\right)+2\gamma_{\!_I}q_{\!_I}^a
v_{\!_Ia}+\pi_{\!_I}^{ab}v_{\!_Ia}v_{\!_Ib}\right\}\,, \label{t5}\\
q_{\!_I}^{*a} &=& q_{\!_I}^a+(\rho_{\!_I}+p_{\!_I})v_{\!_I}^a
\nonumber\\
&&{}+\left\{ (\gamma_{\!_I}-1)q_{\!_I}^a
-\gamma_{\!_I}q_{\!_I}^bv_{\!_Ib}u^a
+\gamma_{\!_I}^2v_{\!_I}^2
\left(\rho_{\!_I}+p_{\!_I}\right)v_{\!_I}^a
+\pi_{\!_I}^{ab}v_{\!_Ib}-\pi_{\!_I}^{bc}v_{\!_Ib}v_{\!_Ic}u^a
\right\} \,,
\label{t6}\\
\pi_{\!_I}^{*ab} &=& \pi_{\!_I}^{ab} +
\left\{-2u^{(a}\pi_{\!_I}^{b)c}v_{\!_Ic}+\pi_{\!_I}^{bc}v_{\!_Ib}
v_{\!_Ic}u^au^b\right\} \nonumber\\
&&{}+\left\{-\case{1}/{3}\pi_{\!_I}^{cd}v_{\!_Ic}v_{\!_Id}h^{ab}+
\gamma_{\!_I}^2\left(\rho_{\!_I}+p_{\!_I}\right)
v_{\!_I}^{\la a}v_{\!_I}^{b\ra}+2\gamma_{\!_I}v_{\!_I}^{\la a}q_{\!_I}
^{b\ra}
\right\}\,.
\label{t7}
\end{eqnarray}
These expressions are the nonlinear generalization of well-known
linearized results (see e.g. \cite{dbe,Is}).
FLRW linearization implies that $v_{\!_I}\ll 1$ for each $I$, and we
neglect all terms which are $O(v_{\!_I}^2)$ or $O(\epsilon v_{\!_I})$.
This removes all terms in braces, dramatically simplifying
the expressions:
\[
\rho_{\!_I}^*\approx\rho_{\!_I}\,,~p_{\!_I}^*\approx
p_{\!_I}\,,~
q_{\!_I}^{*a}\approx q_{\!_I}^a+(\rho_{\!_I}+p_{\!_I})v_{\!_I}^a\,,~
\pi_{\!_I}^{*ab}\approx\pi_{\!_I}^{ab} \,.
\]
To linear order, there is no difference
in the dynamic quantities when
measured in the $I$-frame or the fundamental frame,
apart from a simple velocity correction to the energy flux.
But in the general nonlinear case, this is no longer true.

The total dynamic quantities are simply given by
\[
\rho=\sum_I\rho_{\!_I}^*\,,~p=\sum_I p_{\!_I}^*\,,~
q^{a}=\sum_I q_{\!_I}^{*a}\,,~\pi^{ab}=\sum_I\pi_{\!_I}^{*ab}\,.
\]
Note that
the equations (\ref{t4})--(\ref{t7}) have been
written to make clear the linear parts, so that the irreducible nature
is not explicit. Irreducibility (in the $u^a$-frame) is
revealed on using the relations
\begin{eqnarray*}
q_{\!_I}^{\la a\ra}&\equiv & h^a{}_bq_{\!_I}^b
=q_{\!_I}^a-q_{\!_I}^bv_{\!_Ib}u^a\,,\\
\pi_{\!_I}^{\la a \ra \la  b\ra}&\equiv &h^a{}_ch^b{}_d\pi_{\!_I}^{cd}
=\pi_{\!_I}^{ab}-2u^{(a}\pi_{\!_I}^{b)c}
v_{\!_Ic}+\pi_{\!_I}^{cd}v_{\!_Ic}v_{\!_Id}u^au^b\,.
\end{eqnarray*}

The exact equations show in detail the specific
couplings and contributions
of all partial dynamic quantities in the total quantities.
For example, it is clear that in spatially homogeneous but anisotropic
models, the partial energy fluxes $q_{\!_I}^a$ contribute to the total
energy density, pressure and anisotropic stress at first order in
the velocities $v_{\!_I}$, while the partial anisotropic stresses
$\pi_{\!_I}^{ab}$ contribute to the total energy flux at first order
in $v_{\!_I}$.

The total and partial 4-velocities
define corresponding number 4-currents:
\be
N^a=nu^a+j^a=\sum_IN_{\!_I}^a\,,~N_{\!_I}^a=n_{\!_I}u_{\!_I}^a
+j_{\!_I}^a\,,
\label{t8}\ee
where $n$ and $n_{\!_I}$ are the number densities, $j^a$ and
$j_{\!_I}^a$ are the number fluxes,
and $j_au^a=0=j_{\!_Ia}u_{\!_I}^a$.
It follows that
\begin{eqnarray}
n &=&\sum_In_{\!_I}^*=
\sum_In_{\!_I}+\sum_I\left\{(\gamma_{\!_I}-1)n_{\!_I}+
j_{\!_I}^av_{\!_Ia}\right\} \,, \label{t9}\\
j^a &=&\sum_Ij_{\!_I}^{*a}= \sum_I\left(j_{\!_I}^a
+n_{\!_I}v_{\!_I}^a\right)
+\sum_I\left\{(\gamma_{\!_I}-1)n_{\!_I}v_{\!_I}^a
-v_{\!_I}^bj_{\!_Ib}u^a\right\}\,, \label{t10}
\end{eqnarray}
where the starred quantities are as measured in the $u^a$-frame.
Linearization removes the terms in braces,
regaining the expressions in \cite{dbe,Is}.

Four-velocities may be chosen
in a number of covariant and physical ways. The main choices
are \cite{Ehlers,Is}: (a) the energy (Landau-Lifshitz)
frame, defined by vanishing
energy flux, and (b) the particle (Eckart) frame, defined by vanishing
particle number flux. For a given single fluid, these frames coincide
in equilibrium, but in general they are different.
For each partial $u_{\!_I}^a$,
any change in choice $u_{\!_I}^a\rightarrow \t{u}_{\!_I}^a$
leads to transformations in the partial dynamic quantities,
that are given by equations (\ref{ab7})--(\ref{ab10}) in the appendix.
For the fundamental $u^a$, a change in choice leads
in addition to transformations
of the kinematic quantities, given by
equations (\ref{ab3})--(\ref{ab6}),
and of the gravito-electric/magnetic field, given by
equations (\ref{ab11})--(\ref{ab12}).

A convenient choice for each partial four-velocity $u_{\!_I}^a$
is the energy
frame, i.e. $q_{\!_I}^a=0$ for each $I$ (this is the obvious
choice in the cases $I=C,B$).
As measured in the fundamental frame, the partial energy fluxes do
not vanish, i.e.
$q_{\!_I}^{*a}\neq0$,
and the total energy flux is given by
\begin{equation}
q^a=\sum_I\left[\left(\rho_{\!_I}+p_{\!_I}\right)v_{\!_I}^a+
\pi_{\!_I}^{ab}v_{\!_Ib}+O(\epsilon v_{\!_I}^2,v_{\!_I}^3) \right]\,.
\label{t11}
\end{equation}
With this choice, using the above equations, we find the following
expressions for the dynamic quantities of matter as measured in
the fundamental frame. For
cold dark matter:
\begin{eqnarray}
&& \rho_{\!_C}^*=\gamma_{\!_C}^2\rho_{\!_C}\,,~~p_{\!_C}^*=
{\ts{1\over3}}
\gamma_{\!_C}^2v_{\!_C}^2\rho_{\!_C}\,,\label{t12}\\
&& q_{\!_C}^{*a}=\gamma_{\!_C}^2\rho_{\!_C}v_{\!_C}^a\,,~~
\pi_{\!_C}^{*ab}=\gamma_{\!_C}^2\rho_{\!_C}
v_{\!_C}^{\langle a}v_{\!_C}^{b\rangle}\,. \label{t13}
\end{eqnarray}
For baryonic matter:
\begin{eqnarray}
&& \rho_{\!_B}^*=\gamma_{\!_B}^2\left(1+w_{\!_B}v_{\!_B}^2\right)
\rho_{\!_B}\,,
~~p_{\!_B}^*=\left[w_{\!_B}+{\ts{1\over3}}
\gamma_{\!_B}^2v_{\!_B}^2(1+w_{\!_B})\right]\rho_{\!_B}\,,
\label{t14}\\
&& q_{\!_B}^{*a}=\gamma_{\!_B}^2(1+w_{\!_B})\rho_{\!_B}v_{\!_B}^a\,,~~
\pi_{\!_B}^{*ab}=\gamma_{\!_B}^2(1+w_{\!_B})\rho_{\!_B}
v_{\!_B}^{\langle a}v_{\!_B}^{b\rangle}\,, \label{t15}
\end{eqnarray}
where $w_{\!_B}\equiv p_{\!_B}/\rho_{\!_B}$.
In the case of radiation and neutrinos, we will evaluate the
dynamic quantities relative to the $u^a$-frame directly
via kinetic theory, in the next section.

The total energy-momentum tensor is conserved, i.e.
$\nabla_bT^{ab}=0$, which is equivalent to the evolution
equations (\ref{e1}) and (\ref{e3}).
The partial energy-momentum tensors obey
\begin{equation}
\nabla_bT_{\!_I}^{ab}=J_{\!_I}^{a}=U_{\!_I}^*u^a+M_{\!_I}^{*a}\,,
\label{t16}
\end{equation}
where $U_{\!_I}^*$ is the rate of energy density transfer to
species $I$ as measured in the $u^a$-frame, and
$M_{\!_I}^{*a}=M_{\!_I}^{*\langle a\rangle}$ is the rate
of momentum density transfer to species $I$, as measured in
the $u^a$-frame. Cold dark matter and neutrinos are decoupled
during the period of relevance for CMB anisotropies,
while
radiation and baryons are coupled
through Thomson scattering. Thus
\be
J_{\!_C}^a=0=J_{\!_N}^a\,,~~
J_{\!_R}^a=-J_{\!_B}^a=U_{\!_T}u^a+M_{\!_T}^a\,,
\label{t16a}
\ee
where
the Thomson rates are
\begin{eqnarray}
U_{\!_T}&=&\ne\st\left({\ts{4\over3}}\rho_{\!_R}^*\vb^2-q_{\!_R}^{*a}
v_{\!_Ba}\right)+O(\epsilon\vb^2,\vb^3) \,,
\label{t17}\\
M_{\!_T}^a &=&\ne\st\left({\ts{4\over3}}\rho_{\!_R}^*\vb^a
-q_{\!_R}^{*a}+\pi_{\!_R}^{*ab}v_{\!_Bb}\right)
+O(\epsilon\vb^2,\vb^3) \,,
\label{t18}
\end{eqnarray}
as given by Eq. (\ref{r9a}), derived
in Section V. Here $\ne$ is the free electron number
density, and
$\st$ is the Thomson cross-section.
Note that to linear order, there is no energy transfer,
i.e. $U_{\!_T}\approx 0$.

Using equations (\ref{t12})--(\ref{t15}) in (\ref{t16}), we find
that for cold dark matter
\begin{eqnarray}
\dot{\rho}_{\!_C}+\Theta\rho_{\!_C}+\rho_{\!_C}\div v_{\!_C} &=&
-\left(\rho_{\!_C}v_{\!_C}^2\right)^{\rd}-{\ts{4\over3}}v_{\!_C}^2
\Theta \rho_{\!_C} \nonumber\\
&&{}-v_{\!_C}^a\D_a\rho_{\!_C}-2\rho_{\!_C}A_av_{\!_C}^a+
O(\epsilon v_{\!_C}^2,v_{\!_C}^3) \,, \label{t19}\\
\dot{v}_{\!_C}^a+{\ts{1\over3}}\Theta v_{\!_C}^a+A^a &=&
A_bv_{\!_C}^bu^a-\sigma^a{}_bv_{\!_C}^b \nonumber\\
&&{} +[\omega,v_{\!_C}]^a-v_{\!_C}^b\D_bv_{\!_C}^a+
O(\epsilon v_{\!_C}^2,v_{\!_C}^3) \,, \label{t20}
\end{eqnarray}
and for baryonic matter
\begin{eqnarray}
&&\dot{\rho}_{\!_B}+\Theta(1+w_{\!_B})\rho_{\!_B}+(1+w_{\!_B})
\rho_{\!_B}\div v_{\!_B} \nonumber\\
{}\mbox{~~~}&&{}=-\left[(1+w_{\!_B})\rho_{\!_B}v_{\!_B}^2\right]^{\rd}
-{\ts{4\over3}}v_{\!_B}^2\Theta(1+w_{\!_B})
\rho_{\!_B}
-v_{\!_B}^a\D_a\left[(1+w_{\!_B})\rho_{\!_B}\right] \nonumber\\
{}\mbox{~~~~}&&{}-2(1+w_{\!_B})\rho_{\!_B}A_av_{\!_B}^a
-\ne\st\left({\ts{4\over3}}\rho_{\!_R}^*\vb^2-
q_{\!_R}^{*a}v_{\!_Ba}\right)
+O(\epsilon v_{\!_B}^2,v_{\!_B}^3) \,, \label{t21}\\
&&(1+w_{\!_B})\dot{v}_{\!_B}^a+\left({\ts{1\over3}}-c_{\!_B}^2\right)
\Theta v_{\!_B}^a+(1+w_{\!_B})A^a \nonumber\\
{}\mbox{~~}&&{}+\rho_{\!_B}^{-1}\D^ap_{\!_B}+\rho_{\!_B}^{-1}\ne\st
\left(
\rho_{\!_R}^*\vb^a-q_{\!_R}^{*a}\right)\nonumber\\
{}\mbox{~~~}&&{}=
(1+w_{\!_B})A_bv_{\!_B}^bu^a-(1+w_{\!_B})\sigma^a{}_bv_{\!_B}^b
 +(1+w_{\!_B})[\omega,v_{\!_B}]^a \nonumber\\
 {}\mbox{~~~~}&&{}-(1+w_{\!_B})v_{\!_B}^b\D_bv_{\!_B}^a+
 c_{\!_B}^2(1+w_{\!_B})(\div \vb)\vb^a\nonumber\\
{}\mbox{~~~~}&&{}-\rho_{\!_B}^{-1}\ne\st\pi_{\!_R}^{*ab}v_{\!_Bb}+
O(\epsilon v_{\!_B}^2,v_{\!_B}^3) \,, \label{t22}
\end{eqnarray}
where
$c_{\!_B}^2\equiv\dot{p}_{\!_B}/\dot{\rho}_{\!_B}$
(this equals
the adiabatic sound speed only to linear order).
These conservation equations generalize those given in \cite{cl2}
to the nonlinear case.
FLRW linearization reduces the right hand sides of these
equations to zero, dramatically simplifying the equations.
The conservation equations for the massless
species (radiation and neutrinos) are given below.
Note from Eq. (\ref{t20})
that if the cold dark matter frame
is chosen as the fundamental frame, then the 4-acceleration
vanishes, i.e. $v_{\!_C}^a=0$ implies $A_a=0$.
This is the choice of fundamental frame advocated in \cite{cl2}.

\section{Covariant Lagrangian kinetic theory}

Relativistic kinetic theory (see
e.g. \cite{LQ,Ehlers,Stewart,dGvL,Br})
provides a self-consistent
microscopically based treatment
where there is a natural unifying framework
in which to deal with a gas of particles
in circumstances
ranging from hydrodynamic to free-streaming behavior.
The photon gas undergoes a
transition from hydrodynamic tight coupling with matter,
through the process of
decoupling from matter, to non-hydrodynamic free streaming.
This transition is characterized by the evolution of the
photon mean free path from effectively zero to effectively
infinity. The range of behavior
can appropriately be described by kinetic theory with
Thomson scattering \cite{wey,uza},
and the baryonic matter with which radiation interacts can
reasonably be described hydrodynamically during these times.
(The basic physics of radiation and matter and density perturbations
in cosmology was developed in the works of Sachs and Wolfe \cite{SW},
Silk \cite{sil}, Peebles and Yu \cite{PY},
Weinberg \cite{wei}, and others.)

In the covariant Lagrangian approach
of \cite{ETMa} (see also \cite{EGS,TE}),
the photon 4-momentum $p^a$ (where $p^ap_a=0$) is split as
\begin{equation}
p^a=E(u^a+e^a)\,,~~e^a e_a=1\,,~e^a u_a=0\,,  \label{E}
\end{equation}
where $E=-u_ap^a$ is the energy and $e^a=p^{\la a\ra}/E$
is the direction, as
measured by a comoving (fundamental)
observer. Then the photon
distribution function
is decomposed into covariant harmonics
via the expansion \cite{ETMa,T2}
\begin{equation}
f(x,p)=f(x,E,e)
=F+F_ae^a+F_{ab}e^ae^b+\cdots\,= \sum_{\ell\geq0} F_{A_l}(x,E)
e^{\la A_l\ra},
\label{r3}\end{equation}
where ${e}^{A_\ell}\equiv e^{a_1}e^{a_2}\cdots e^{a_\ell}$, and
$e^{\la A_\ell\ra}$
provides a representation of the rotation group \cite{GE}.
The covariant multipoles are
irreducible since they
are PSTF, i.e.
\[
F_{a\cdots b}=F_{\la a\cdots b\ra}~~\Leftrightarrow~~
F_{a\cdots b}=F_{(a\cdots b)}\,,~F_{a\cdots b}u^b=0=
F_{a\cdots bc}h^{bc}\,.
\]
They encode the anisotropy structure of the distribution in the
same way as the usual spherical harmonic expansion
\[
f=\sum_{\ell\geq0}\,
\sum_{m=-\ell}^{+\ell}f_\ell^m(x,E)Y_\ell^m(\vec{e}\,)
\,,
\]
but here (a) the $F_{A_\ell}$ are covariant, and thus independent
of any choice of coordinates in momentum space, unlike the
$f_\ell^m$; (b) $F_{A_\ell}$ is a rank-$\ell$ tensor field
on spacetime
for each fixed $E$, and directly determines the $\ell$-multipole
of radiation anisotropy after integration over $E$.
The multipoles can be recovered from the distribution function
via \cite{ETMa,GE}
\be
F_{A_\ell}= \Delta_\ell^{-1}
\int f(x,E,e) e_{\la A_\ell\ra}\d\Omega\,,
~\mbox{ with }~\Delta_{\ell} = 4\pi
{(\ell!)^2 2^{\ell} \over (2 \ell+1)!}\,,
\label{r6}\ee
where $\d\Omega=\d^2{e}$ is
a solid angle in momentum space. A further useful identity is
\cite{ETMa}
\be
\int {e}^{A_{\ell}}\d\Omega=
{4\pi\over \ell+1}\left\{
\begin{array}{ll}
0 & \ell ~\mbox{ odd}\,, \\
{}&{} \\
h^{(a_1a_2}
h^{a_3a_4}\cdots h^{a_{\ell-1}a_\ell)} & \ell ~\mbox{ even} \,.
\end{array}
\right.
\label{r7}\ee

The first 3 multipoles arise from
the radiation energy-momentum tensor, which is
\[
T_{\!_R}^{ab}(x)=\int p^ap^bf(x,p)\d^3p=
\rho_{\!_R}^*u^au^b+{\ts{1\over3}}\rho_{\!_R}^*h^{ab}
+2q_{\!_R}^{*(a}u^{b)}+\pi_{\!_R}^{*ab}\,,
\]
where
$\d^3p=E\d E\d\Omega$ is the covariant volume element
on the future null cone at event $x$.
It follows that the dynamic quantities of the radiation
(in the $u^a$-frame) are:
\be
\rho_{\!_R}^* = 4\pi\int_0^\infty E^3F\,\d E\,,
~q_{\!_R}^{*a} = {4\pi\over 3}\int_0^\infty E^3F^a\,\d E \,,
~\pi_{\!_R}^{*ab} = {8\pi\over 15}\int_0^\infty E^3F^{ab}\,\d E\,.
\label{em3}
\ee
From now on, we
\underline{drop the asterisks} from the radiation
dynamic quantities relative to the fundamental frame,
since we do not need to relate them to their values
in the radiation frame.

We extend these dynamic quantities
to all multipole orders by defining\footnote{
Because photons are
massless, we do not need the complexity of the
moment
definitions used in \cite{ETMa}.
In \cite{cl2}, $J_{A_\ell}^{(\ell)}$ is used, where
$J_{A_\ell}^{(\ell)}=\Delta_\ell \Pi_{A_\ell}$.
From now on, all energy integrals will be understood
to be over the range $0\leq E\leq\infty$.
}
\cite{ETMa}
\be
\Pi_{a_1\cdots a_\ell} =
\int E^3 F_{a_1\cdots a_\ell}\d E\,,
\label{r10}\ee
so that
$\Pi=\rho_{\!_R}/4\pi$, $\Pi^a=3q_{\!_R}^a/4\pi$ and
$\Pi^{ab}=15\pi_{\!_R}^{ab}/8\pi$.

The Boltzmann equation is
\begin{equation}
{\d f\over \d v}\equiv p^a{\p f\over \p x^a}-\Gamma^a{}_{bc}
p^bp^c{\p f\over \p p^a}=C[f] \,,
\label{boltz}
\end{equation}
where $p^a=\d x^a/\d v$ and $C[f]$
is the collision term, which determines the rate of change
of $f$ due to emission, absorption and scattering processes.
This term is also decomposed into covariant harmonics:
\be
C[f]=\sum_{\ell\geq 0} b_{A_\ell}(x,E){e}^{A_\ell}
=b+b_ae^a+b_{ab}e^ae^b+\cdots
\,, \label{scatt}
\ee
where the multipoles $b_{A_\ell}=b_{\la A_\ell \ra}$
encode covariant irreducible properties of the particle interactions.
Then the Boltzmann equation is equivalent to an infinite
hierarchy of covariant multipole equations
\[
L_{A_\ell}(x,E)=b_{A_\ell}[F_{A_m}](x,E)\,,
\]
where $L_{A_\ell}$ are the multipoles of $\d f/\d v$, and
will be given in the next section.
These multipole equations are tensor field equations on spacetime for
each
value of the photon energy $E$ (but note that energy changes along
each
photon path).  Given the solutions $F_{A_\ell}(x,E)$ of the
equations, the relation (\ref{r3}) then determines the full photon
distribution $f(x,E,e)$ as a scalar field over phase space.

Over the period of importance for CMB anisotropies, i.e. considerably
after electron-positron annihilation, the average photon energy
is much less than the electron rest mass and the electron thermal
energy
may be neglected, so that the Compton interaction between
photons and electrons (the dominant interaction between
radiation and matter) may reasonably be described in the Thomson
limit. (See \cite{pl} for refinements.)
We will also neglect the effects of polarization (see e.g.
\cite{zs}). For Thomson scattering
\be
C[f]=\st \ne E_{\!_B}\left[\bar{f}(x,p)-f(x,p)\right]\,,
\label{r1}\ee
where
$E_{\!_B}=-p_a\ub^a$ is the
photon energy
relative to the baryonic (i.e. baryon-electron)
frame $\ub^a$, and $\bar{f}(x,p)$
determines the number of photons scattered into the phase space
volume element at $(x,p)$.
The differential Thomson cross-section is
proportional to $1+\cos^2\alpha$, where $\alpha$ is the angle between
initial and final photon directions in the baryonic frame.
Thus $\cos\alpha=\eb^ae'_{\!_Ba}$
where $e'_{\!_Ba}$ is the initial and $\eb^a$ is the final
direction, so that
\[
p'^a=E_{\!_B}\left(\ub^a+\eb'^a\right)\,,~
p^a=E_{\!_B}\left(\ub^a+\eb^a\right)\,,
\]
where we have used $E'_{\!_B}=E_{\!_B}$, which follows since the
scattering is elastic. Here
$\ub^a$ is given by Eq. (\ref{t3}),
where $\vb^a$ is the velocity
of the baryonic frame relative to the fundamental frame $u^a$,
with $\vb^au_a=0$.
Then
$\bar{f}$ is given by \cite{cl2,pl}
\be
\bar{f}(x,p)={3\over 16\pi}\int f(x,p')\left[1+\left(
\eb^{a}e'_{\!_Ba}\right)^2\right]\d\Omega'_{\!_B} \,.
\label{r4}\ee
The exact forms of the photon energy and direction
in the baryonic frame follow on
using equations (\ref{t3}) and (\ref{ab1}):
\begin{eqnarray}
E_{\!_B} &=& E \gamma_{\!_B}\left(1-\vb^a e_a\right) \,, \label{r22}\\
\eb^a &=& {1\over \gamma_{\!_B}(1-\vb^ce_c)}\left[e^a+
\gamma^2_{\!_B}\left(\vb^be_b-\vb^2\right)u^a+\gamma^2_{\!_B}
\left(\vb^be_b-1\right)\vb^a\right]\,. \label{r23}
\end{eqnarray}

Anisotropic scattering will source polarization, and
small errors are introduced by assuming that the radiation remains
unpolarized \cite{hssw}. A fully consistent and general treatment
requires
the incorporation of polarization. However, for simplicity, and
in line with many previous treatments, we will neglect polarization
effects.

\section{The nonlinear multipole hierarchy}

The full Boltzmann equation in photon phase space contains more
information than necessary to analyze radiation anisotropies
in an inhomogeneous universe. For that purpose,
  when the radiation is close to black-body we do not require
the full spectral behaviour of the distribution multipoles, but
only the energy-integrated multipoles. The monopole
leads to the average temperature,  while the higher order multipoles
determine the temperature fluctuations. The 1+3 covariant and
gauge-invariant
definition of the average temperature $T$ is given by \cite{mes}
\begin{equation}
\rho_{\!_R}(x)=4\pi\int E^3F(x,E)\d E=rT(x)^4\,,
\label{r27}
\end{equation}
where $r$ is the radiation constant. If $f$ is close to a Planck
distribution, then $T$ is the thermal black-body average
temperature. But note
that no notion of {\em background} temperature is involved in this
definition. There is an all-sky average implied in (\ref{r27}).
Fluctuations across the sky are measured by integrating the
higher multipoles (a precise definition is given below), i.e.
the fluctuations are determined by the
$\Pi_{a_1\cdots a_\ell}$ ($\ell\geq1$)
defined in Eq. (\ref{r10}).

The form of $C[f]$ shows that covariant equations for the temperature
fluctuations arise from decomposing the energy-integrated Boltzmann
equation
\be
\int E^2{\d f\over \d v}\,\d E=\int E^2 C[f]\d E \label{ibe}
\ee
into 1+3 covariant multipoles. We begin with the right
hand side, which
requires the covariant form of the Thomson scattering term (\ref{r4}).
Since the baryonic frame will
move nonrelativistically relative to the fundamental frame in
all cases of physical interest, it is sufficient to
linearize only in $\vb$, and not in the other quantities.
Thus we drop terms in $O(\epsilon\vb^2,\vb^3)$ but do not neglect
terms that are $O(\epsilon^0\vb^2,\epsilon\vb)$
or $O(\epsilon^2)$ relative to the FLRW limiting background.
In other words, we make no restrictions on the
geometric and physical quantities that covariantly
characterize the spacetime, apart from assuming a
nonrelativistic relative average velocity for matter. The resulting
expression will in particular be applicable for covariant
second-order effects in FLRW backgrounds (recognising that
polarization effects should be included for a complete treatment),
or for first-order effects
in Bianchi backgrounds.

For brevity, we will use the notation
\[
{\cal O}[3]\equiv O(\epsilon\vb^2,\vb^3)\,,
\]
noting that this does {\em not}
imply any second-order restriction
on the dynamic, kinematic and gravito-electric/magnetic quantities.
It follows from equations (\ref{r7}) and (\ref{r4}) that
\be
4\pi\int \bar{f}E_{\!_B}^3\d E_{\!_B}=(\rho_{\!_R})_{\!_B}+
{\ts{3\over4}}(\pi_{\!_R}^{ab})_{\!_B}e_{\!_Ba}e_{\!_Bb} \,,
\label{r27a}
\ee
where the dynamic radiation quantities are evaluated in the
{\em baryonic} frame. This approach relies on the
frame-transformations given in the appendix, and allows us to
evaluate the Thomson scattering integral more directly and
clearly than other approaches. In the process,
we are also generalizing to
include nonlinear effects. We use equations (\ref{ab7}) and
(\ref{ab10}) to transform back to the
fundamental frame:\footnote{
As noted in Section III, we
retain the $O(\vb^2)$ term in $(\rho_{\!_R})_{\!_B}$ since
$\rho_{\!_R}$ is zero-order.
}
\begin{eqnarray*}
(\rho_{\!_R})_{\!_B} &=& \rho_{\!_R}\left[1+{\ts{4\over3}}\vb^2\right]
-2q_{\!_R}^a v_{\!_Ba} +{\cal O}[3]\,,\\
(\pi_{\!_R}^{ab})_{\!_B}& =& \pi_{\!_R}^{ab}+2v_{\!_Bc}
\pi_{\!_R}^{c(a}u^{b)}-2q_{\!_R}^{\la a}\vb^{b\ra}
+{\textstyle{4\over3}}\rho_{\!_R}\vb^{\la a}\vb^{b\ra}
+{\cal O}[3]\,.
\end{eqnarray*}
Now
\begin{eqnarray}
\int E^2C[f]\d E &=& \ne\st\left[1+3\vb^ce_c+\left(\vb^ce_c\right)^2
-{\textstyle{3\over2}}\vb^2\right]
\int E_{\!_B}^3\bar{f}\d E_{\!_B} \nonumber\\
{}&&
-\ne\st\left[1-\vb^ce_c+{\textstyle{1\over2}}\vb^2\right]
\int fE^3 \d E
+{\cal O}[3]\,.
\label{r27b}\end{eqnarray}
In addition, we need the following identity, valid for any projected
vector $v^a$:
\begin{eqnarray}
v^ae_af &=& {\textstyle{1\over3}}F_av^a+
\left[Fv_a+
 {\textstyle{2\over5}}F_{ab}v^b\right]e^a \nonumber\\
 {}&&+\left[F_{\la a}v_{b\ra}+{\textstyle{3\over7}}F_{abc}v^c\right]
 e^{\la a}e^{b\ra}+\cdots \nonumber\\
 {}&=&\sum_{\ell\geq0}\left[F_{\la A_{\ell-1}}v_{a_\ell\ra}+
 \left({\ell+1 \over 2\ell+3}\right)F_{A_\ell a}v^a\right]
 {e}^{\la A_\ell\ra}  \,.
\label{r9}\end{eqnarray}
(Here and subsequently,
we use the convention that $F_{A_\ell}=0$ for $\ell<0$.)
This identity may be proved using Eq.
(\ref{r7}) and the identity
(see \cite{ETMa}, p. 470):
\begin{equation}
V_{\la b}S_{A_{\ell}\ra}=V_{(b}S_{A_{\ell})}
-\left({\ell \over 2 \ell +1}
\right) V^cS_{c(A_{\ell-1}}h_{a_\ell b )}~
\mbox{ where }~S_{A_\ell}=S_{\la A_{\ell} \ra}\,.
\label{r20}\end{equation}
Using the above equations, we find that\footnote{
A. Challinor has independently derived the same result
\cite{c}.
}
\begin{eqnarray}
 4\pi\int E^2C[f]\d E &= &
\ne\st\left[ {\textstyle{4\over3}}\rho_{\!_R}\vb^2-
q_{\!_R}^av_{\!_Ba}\right] \nonumber\\
&&{}- \ne\st\left[3q_{\!_R}^a-4\rho_{\!_R}\vb^a-
3\pi_{\!_R}^{ab}v_{\!_Bb}\right]e_a \nonumber\\
&&{}
-\ne\st\left[{\ts{27\over4}}\pi_{\!_R}^{ab}-{\ts{3\over2}}
q_{\!_R}^{\la a}\vb^{b\ra}-{\ts{12\over7}}\pi
\Pi^{abc}v_{\!_Bc}
-3\rho_{\!_R}\vb^{\la a}\vb^{b\ra}
\right]e_{\la a}e_{b\ra} \nonumber\\
&&{}-\ne\st\left[4\pi\Pi^{abc}-{\ts{45\over4}}
\pi_{\!_R}^{\la ab}\vb^{c\ra}-{\ts{16\over9}}\pi
\Pi^{abcd}v_{\!_Bd}\right]e_{\la a}e_be_{c\ra}+\cdots
+{\cal O}[3] \,.
\label{r9a}
\end{eqnarray}
Now it is clear from equations (\ref{r27a})
and (\ref{r27b}) that the first four multipoles
are affected by Thomson scattering differently
than the higher multipoles. This is confirmed by the
form of equation (\ref{r9a}). Defining the
energy-integrated scattering multipoles
\[
K_{A_\ell}=\int E^2b_{A_\ell}\d E\,,
\]
we find from Eq. (\ref{r9a}) that
\begin{eqnarray}
K &=& \ne\st\left[{\ts{4\over3}}\Pi\vb^2-
{\ts{1\over3}}\Pi^av_{\!_Ba}\right]+
{\cal O}[3]\,,
\label{rr1}\\
K^a &=&-\ne\st\left[\Pi^a-4\Pi\vb^a-{\ts{2\over5}}\Pi^{ab}v_{\!_Bb}
\right]+{\cal O}[3]\,,
\label{rr2}\\
K^{ab} &=&-\ne\st\left[{\ts{9\over10}}\Pi^{ab}-{\ts{1\over2}}
\Pi^{\la a}\vb^{b\ra}-{\ts{3\over7}}\Pi^{abc}v_{\!_Bc}
-3\Pi\vb^{\la a}\vb^{b\ra}
\right]+{\cal O}[3]\,,
\label{rr3}\\
K^{abc} &=&-\ne\st\left[\Pi^{abc}-{\ts{3\over2}}
\Pi^{\la ab}\vb^{c\ra}-{\ts{4\over9}}\Pi^{abcd}v_{\!_Bd}
\right]+{\cal O}[3]\,,
\label{rr3a}
\end{eqnarray}
and, for $\ell> 3$:
\begin{equation}
K^{A_\ell}=
-\ne\st\left[
\Pi^{A_\ell}-\Pi^{\la A_{\ell-1}} \vb^{a_\ell\ra}
-\left({\ell+1\over 2\ell+3}\right)
\Pi^{A_\ell a}v_{\!_Ba}\right]
+{\cal O}[3]\,.
\label{r21}
\end{equation}

Equations (\ref{rr1})--(\ref{r21})
are a nonlinear generalization of
the results given by Challinor and Lasenby \cite{cl2}.
They show the new
{\em coupling of baryonic bulk velocity to the
radiation multipoles, arising from local nonlinear
effects in Thomson scattering.}
If we linearize fully, i.e. neglect all terms containing $\vb$
except the $\rho_{\!_R}\vb^a$ term in the dipole
$K^a$, which is first-order,
then our equations reduce to those in \cite{cl2}.
The generalized nonlinear equations apply to the
analysis of second-order effects on an FLRW
background, to first-order effects on a spatially homogeneous
but anisotropic background, and more generally, to any situation
where the baryonic frame is non-relativistic relative to the
fundamental $u^a$-frame.

Next we require the multipoles of $\d f/\d v$.
These can be read directly from the general expressions first
derived in \cite{ETMa}, which are exact, 1+3 covariant and also
include the case of massive particles. For clarity and completeness,
we outline an alternative,  1+3 covariant derivation
(the derivation in \cite{ETMa} uses tetrads). We require
the identity \cite{EGS,TE}
\be
{\d E\over\d v}=-E^2\left[
{\ts{1\over3}}\Theta+A_ae^a+\sigma_{ab}
e^ae^b\right]\,, \label{dEdv}
\ee
which follows directly from $E=-p^au_a$,
$p^b\nabla_bp^a=0$ and $\nabla_bu_a=-A_au_b+\D_bu_a$.
Then
\begin{eqnarray*}
{\d\over\d v}\left[F_{a_1\cdots a_\ell}(x,E)e^{a_1}\cdots
e^{a_{\ell}}\right] &=&
{\d\over\d v}\left[E^{-\ell}F_{a_1\cdots a_\ell}(x,E)p^{a_1}\cdots
p^{a_{\ell}}\right] \\
{}&=& E\left\{
\left[{\ts{1\over3}}
\Theta+A_be^b+\sigma_{bc}e^be^c\right]\left(\ell
F_{a_1\cdots a_\ell}-EF'_{a_1\cdots a_\ell}\right)
e^{a_1}\cdots{e}^{a_\ell}
\right. \\
&&{}\left.
+\left(u^{a_1}+e^{a_1}\right)\cdots\left(u^{a_\ell}+
e^{a_\ell}\right)\left[\dot{F}_{a_1\cdots a_\ell}
+e^b\nabla_bF_{a_1\cdots a_\ell}
\right]\right\}\,,
\end{eqnarray*}
where a prime denotes $\partial/\partial E$.
The first term is readily put into irreducible PSTF form using
the identity (\ref{r20}) with $V_a=A_a$,
and its extension to the case when $V_a$
is replaced by a rank-2 PSTF tensor $W_{ab}$ (see
\cite{ETMa}, p. 470), with $W_{ab}=\sigma_{ab}$.
In the second term, when the round brackets are expanded, only
those terms with at most one $u^{a_r}$
survive, and
\[
u^{a} \dot{F}_{a\cdots}=- A^{a} F_{a\cdots}\,,~~
u^{b} \nabla^{a} F_{b\cdots}=
-\left({\ts{1\over3}}\Theta h^{ab}+
\sigma^{ab}-\ep^{abc}\omega_c \right)F_{b\cdots}\,.
\]
Thus the covariant multipoles $b_{A_\ell}$ of $\d f/\d v$ are
\begin{eqnarray}
E^{-1}b_{A_\ell} &=&
\dot{F}_{\la A_\ell \ra}-{\ts{1\over3}}\Theta EF'_{A_\ell}
+\D_{\la a_\ell} F_{A_{\ell-1}\ra} +{(\ell+1)\over(2\ell+3)}
\D^aF_{aA_\ell}
\nonumber\\
&&{}
-{(\ell+1)\over(2\ell+3)}E^{-(\ell+1)}\left[E^{\ell+2}F_{aA_\ell}
\right]'A^a-E^\ell\left[E^{1-\ell}F_{\la A_{\ell-1}}\right]'
A_{a_\ell\ra}
\nonumber\\
&&{}
-\ell\omega^b\ep_{bc( a_\ell}F_{A_{\ell-1})}{}^c
-{(\ell+1)(\ell+2)\over(2\ell+3)(2\ell+5)}E^{-(\ell+2)}
\left[E^{\ell+3}F_{abA_\ell}\right]'\sigma^{ab}
\nonumber\\
&&{}
-{2\ell\over (2\ell+3)}E^{-1/2}\left[E^{3/2}F_{b\la A_{\ell-1}}
\right]'\sigma_{a_\ell\ra}{}^b-E^{\ell-1}\left[E^{2-\ell}
F_{\la A_{\ell-2}}\right]'\sigma_{a_{\ell-1}a_\ell\ra}\,.
\label{r25}\end{eqnarray}

This regains the result of \cite{ETMa} [equation (4.12)] in the
massless case, with minor corrections. The form given here
benefits from the streamlined version of the 1+3 covariant formalism.
We reiterate that this result is exact and holds for any
photon or (massless) neutrino distribution in any spacetime.
We now multiply Eq. (\ref{r25}) by $E^3$ and
integrate over all energies, using integration by parts
and the fact that $E^nF_{a\cdots}\rightarrow0$
as $E\rightarrow\infty$ for any positive $n$. We
obtain the multipole equations that determine the
brightness multipoles $\Pi_{A_{\ell}}$:
\begin{eqnarray}
K_{A_\ell} &=&
\dot{\Pi}_{\la A_\ell\ra}+{\ts{4\over3}}\Theta \Pi_{A_\ell}+
\D_{\la a_\ell}\Pi_{A_{\ell-1}\ra}
+{(\ell+1)\over(2\ell+3)}\D^b \Pi_{bA_\ell}
\nonumber\\
&&{}
-{(\ell+1)(\ell-2)\over(2\ell+3)} A^b \Pi_{bA_\ell}
+(\ell+3) A_{\la a_\ell} \Pi_{A_{\ell-1}\ra}
-\ell\omega^b\ep_{bc( a_\ell} \Pi_{A_{\ell-1})}{}^c
\nonumber\\
&&{}
-{(\ell-1)(\ell+1)(\ell+2)\over(2\ell+3)(2\ell+5)}
\sigma^{bc}\Pi_{bcA_\ell}+{5\ell\over(2\ell+3)}
\sigma^b{}_{\la a_\ell} \Pi_{A_{\ell-1}\ra b}
-(\ell+2)
\sigma_{\la a_{\ell}a_{\ell-1}} \Pi_{A_{\ell-2}\ra}\,.
\label{r26}\end{eqnarray}

Once again, this is an exact result, and it holds also
for any collision term, i.e. any $K_{A_\ell}$.
For decoupled neutrinos, we have $K_{\!_N}^{A_\ell}=0$
in this equation.
For photons undergoing Thomson scattering,
the left hand side of Eq. (\ref{r26}) is given by Eq. (\ref{r21}),
which is exact in the kinematic and dynamic quantities, but first
order in the relative baryonic velocity. The equations (\ref{r21})
and (\ref{r26}) thus constitute a nonlinear generalization of the
FLRW-linearized case given by Challinor and Lasenby \cite{cl2}.

These equations describe evolution along
the timelike world-lines of fundamental observers,
not along the lightlike geodesics of photon motion.
The timelike integration
is related to light cone integrations by making homogeneity
assumptions about the distribution of matter in (spacelike) surfaces
of constant time, as is discussed in \cite{gd}.

The monopole and dipole of equation (\ref{r26}) give the
evolution equations of energy and momentum density:
\begin{eqnarray}
K &=& \dot{\Pi}+{\ts{4\over3}}\Theta\Pi+{\ts{1\over3}}\D^a\Pi_a
+{\ts{2\over3}}A^a\Pi_a+{\ts{2\over15}}\sigma^{ab}\Pi_{ab}
\,,\label{r26a}\\
K^a &=& \dot{\Pi}^{\la a\ra}+{\ts{4\over3}}\Theta\Pi^a+\D^a\Pi
+{\ts{2\over5}}\D_b\Pi^{ab}
\nonumber\\
{}&&+{\ts{2\over5}}A_b\Pi^{ab}+4\Pi A^a-[\omega,\Pi]^a+
\sigma^{ab}\Pi_b \,.
\label{r26b}
\end{eqnarray}
In the case of neutrinos,
$K_{\!_N}=0=K_{\!_N}^{a}$, these express
the conservation of energy and momentum:\footnote{
As in the photon case,
we omit the asterisks on the neutrino dynamic quantities,
since we do not require their values in the neutrino frame.
}
\begin{eqnarray}
\dot{\rho}_{\!_N} +{\ts{4\over3}}\Theta\rho_{\!_N}+\D_a q_{\!_N}^a
&=& -2A_aq_{\!_N}^a-\sigma_{ab}\pi_{\!_N}^{ab}
\,,\label{r26c}\\
\dot{q}_{\!_N}^{\la a\ra}+{\ts{4\over3}}\Theta q_{\!_N}^a
+{\ts{4\over3}}\rho_{\!_N}A^a &&
\nonumber\\
+{\ts{1\over3}}\D^a\rho_{\!_N}
+\D_b\pi_{\!_N}^{ab} &=&-[\omega,q_{\!_N}]^a-\sigma^a{}_bq_{\!_N}^b-
A_b\pi_{\!_N}^{ab}\,.
\label{r26d}
\end{eqnarray}
FLRW-linearization reduces the right hand sides to zero.
For photons,
$K$ and $K^a$ are given by equations (\ref{rr1}) and (\ref{rr2}),
and determine the Thomson rates of transfer in equations
(\ref{t17}) and (\ref{t18}):
\be
U_{\!_T}=4\pi K\,,~~M_{\!_T}^a={4\pi\over3}K^a\,.
\label{r26e}
\ee

Finally, we return to the definition of temperature anisotropies.
As noted above, these are determined by the $\Pi_{A_\ell}$.
Generalizing the linearized 1+3 covariant approach in
\cite{mes}, we define the temperature fluctuation $\tau(x,e)$
via the
directional bolometric
brightness:
\begin{equation}
T(x)\left[1+\tau(x,e)\right]=\left[{4\pi\over r}\int E^3
f(x,E,e)\d E\right]^{1/4}\,.
\label{r28}\end{equation}
This is a 1+3 covariant and gauge-invariant definition which
is also exact. We can rewrite it explicitly in terms of
the $\Pi_{A_\ell}$:
\begin{equation}
\tau(x,e)=\left[1+\left({4 \pi \over\rho_{\!_R}}\right)
\sum_{\ell\geq 1}
\Pi_{A_\ell}{e}^{A_\ell}\right]^{1/4}-1
=\tau_ae^a+\tau_{ab}e^ae^b+\cdots
\,.
\label{r29}\end{equation}
In principle, we can extract the irreducible PSTF
temperature fluctuation multipoles by using the inversion
in Eq. (\ref{r6}):
\begin{equation}
\tau_{A_\ell}(x)={\Delta_{\ell}^{-1} }\int\tau(x,e)
e_{\la A_\ell\ra}\d \Omega\,.
\label{r30}\end{equation}

In the almost-FLRW case, when $\tau$ is $O(\epsilon)$,
we regain from Eq. (\ref{r29}) the linearized definition
given in \cite{mes}:
\begin{equation}
\tau_{A_\ell} \approx \left({\pi \over \rho_{\!_R}}\right)
\Pi_{A_\ell}
\,,
\label{r31}\end{equation}
where $\ell\geq 1$. In particular, the dipole and quadrupole are
\be
\tau^a \approx {3q_{\!_R}^a\over 4\rho_{\!_R}}\,~~\mbox{and}~~
\tau^{ab} \approx {15\pi_{\!_R}^{ab}\over 2\rho_{\!_R}}\,.
\label{dip-quad}
\ee

\section{Qualitative implications of the nonlinear dynamical effects}

In Section II, we gave the nonlinear evolution and constraint
equations governing the kinematic, total dynamic and
gravito-electric/magnetic quantities -- see
equations (\ref{e1})--(\ref{c5}). In these equations,
the total dynamic
quantities are, using the results of Section III:
\begin{eqnarray}
\rho &=& \rho_{\!_R}+\rho_{\!_N}+\left(1+v_{\!_C}^2\right)\rho_{\!_C}
+\left[1+(1+w_{\!_B})\vb^2\right]\rho_{\!_B}
+\Lambda
+{\cal O}[3] \,,\label{nl1}\\
p &=& {\ts{1\over3}}\rho_{\!_R}+{\ts{1\over3}}\rho_{\!_N}+
{\ts{1\over3}}v_{\!_C}^2\rho_{\!_C}+\left[w_{\!_B}+{\ts{1\over3}}
(1+w_{\!_B})\vb^2\right]\rho_{\!_B}
-\Lambda
+{\cal O}[3]\,,\label{nl2}\\
q^a &=& q_{\!_R}^a+q_{\!_N}^a+\rho_{\!_C}^av_{\!_C}^a
+(1+w_{\!_B})\rho_{\!_B}\vb^a
+{\cal O}[3]
\,,\label{nl3}\\
\pi^{ab} &=& \pi_{\!_R}^{ab}+\pi_{\!_N}^{ab}+\rho_{\!_C}v_{\!_C}^
{\la a}
v_{\!_C}^{b\ra}+(1+w_{\!_B})\rho_{\!_B}\vb^{\la a}\vb^{b\ra}
+{\cal O}[3]
\,.\label{nl4}
\end{eqnarray}
The conservation equations for matter were given in Section III
-- see equations (\ref{t19})--(\ref{t22}). For
neutrinos, the equations were given in Section V -- see
equations (\ref{r26c}) and (\ref{r26d}). For photons, the
equations follow from the results of Section V as:
\begin{eqnarray}
\dot{\rho}_{\!_R}+{\ts{4\over3}}\Theta\rho_{\!_R}+\D_a q_{\!_R}^a
+2A_aq_{\!_R}^a+\sigma_{ab}\pi_{\!_R}^{ab}
&=&\ne\st\left({\ts{4\over3}}\rho_{\!_R}\vb^2-q_{\!_R}^av_{\!_Ba}
\right)+{\cal O}[3]\,,
\label{nl5}\\
\dot{q}_{\!_R}^{\la a\ra}+{\ts{4\over3}}\Theta q_{\!_R}^a
+{\ts{4\over3}}\rho_{\!_R}A^a+{\ts{1\over3}}\D^a\rho_{\!_R}
+\D_b\pi_{\!_R}^{ab}
&&\nonumber\\
+\sigma^a{}_bq_{\!_R}^b-[\omega,q_{\!_R}]^a+A_b\pi_{\!_R}^{ab}
&=&\ne\st\left({\ts{4\over3}}\rho_{\!_R}\vb^a-q_{\!_R}^a+
\pi_{\!_R}^{ab}v_{\!_Bb}\right)+{\cal O}[3]\,.
\label{nl6}
\end{eqnarray}
The nonlinear dynamical equations are completed by the integrated
Boltzmann multipole equations given in Section V -- see
Eq. (\ref{r26}).
For neutrinos ($\ell\geq2$):
\begin{eqnarray}
0 &=&
\dot{\Pi}_{\!_N}^{\la A_\ell\ra}+{\ts{4\over3}}\Theta \Pi_{\!_N}
^{A_\ell}+
\D^{\la a_\ell}\Pi_{\!_N}^{A_{\ell-1}\ra}
+{(\ell+1)\over(2\ell+3)}\D_b \Pi_{\!_N}^{bA_\ell}
\nonumber\\
&&{}
-{(\ell+1)(\ell-2)\over(2\ell+3)} A_b \Pi_{\!_N}^{bA_\ell}
+(\ell+3) A^{\la a_\ell} \Pi_{\!_N}^{A_{\ell-1}\ra}
-\ell\omega_b\ep^{bc( a_\ell} \Pi_{\!_N}^{A_{\ell-1})}{}^c
\nonumber\\
&&{}
-{(\ell-1)(\ell+1)(\ell+2)\over(2\ell+3)(2\ell+5)}
\sigma_{bc}\Pi_{\!_N}^{bcA_\ell}+{5\ell\over(2\ell+3)}
\sigma_b{}^{\la a_\ell} \Pi_{\!_N}^{A_{\ell-1}\ra b}
-(\ell+2)
\sigma^{\la a_{\ell}a_{\ell-1}} \Pi_{\!_N}^{A_{\ell-2}\ra}\,.
\label{nl7}\end{eqnarray}
For photons, the quadrupole evolution equation is
\begin{eqnarray}
&&\dot{\pi}_{\!_R}^{\la ab\ra}+{\ts{4\over3}}\Theta \pi_{\!_R}^{ab}
+{\ts{8\over15}}\rho_{\!_R}\sigma^{ab}+
{\ts{2\over5}}\D^{\la a}q_{\!_R}^{b\ra}
+{8\pi\over35}\D_c \Pi^{abc}
\nonumber\\
&&{}+2 A^{\la a} q_{\!_R}^{b\ra}
-2\omega^c\ep_{cd}{}{}^{(a} \pi_{\!_R}^{b) d}
+{\ts{2\over7}}\sigma_c{}^{\la a}\pi_{\!_R}^{b\ra c}
-{32\pi\over315}
\sigma_{cd} \Pi^{abcd}
\nonumber\\
{}&&{}=
-\ne\st\left[{\ts{9\over10}}\pi_{\!_R}^{ab}-{\ts{1\over5}}
q_{\!_R}^{\la a}\vb^{b\ra}-{8\pi\over35}\Pi^{abc}v_{\!_Bc}
-{\ts{2\over5}}\rho_{\!_R}\vb^{\la a}\vb^{b\ra}
\right]+{\cal O}[3]
\,.
\label{nl8}
\end{eqnarray}
In the free-streaming case $\ne=0$, equation (\ref{nl8})
reduces to the result first given in \cite{sme}. This quadrupole
evolution equation is central to the proof that almost-isotropy
of the CMB after last scattering implies almost-homogeneity
of the universe \cite{sme}.

The higher multipoles ($\ell>3$) evolve according to
\begin{eqnarray}
&&\dot{\Pi}^{\la A_\ell\ra}+{\ts{4\over3}}\Theta \Pi^{A_\ell}+
\D^{\la a_\ell}\Pi^{A_{\ell-1}\ra}
+{(\ell+1)\over(2\ell+3)}\D_b \Pi^{bA_\ell}
\nonumber\\
&&{}-{(\ell+1)(\ell-2)\over(2\ell+3)} A_b \Pi^{bA_\ell}
+(\ell+3) A^{\la a_\ell} \Pi^{A_{\ell-1}\ra}
-\ell\omega^b\ep_{bc}{}{}^{( a_\ell} \Pi^{A_{\ell-1}) c }
\nonumber\\
&&{}-{(\ell-1)(\ell+1)(\ell+2)\over(2\ell+3)(2\ell+5)}
\sigma_{bc}\Pi^{bcA_\ell}+{5\ell\over(2\ell+3)}
\sigma_b{}^{\la a_\ell} \Pi^{A_{\ell-1}\ra b}
-(\ell+2)
\sigma^{\la a_{\ell}a_{\ell-1}} \Pi^{A_{\ell-2}\ra}
\nonumber\\
&&{}=
-\ne\st\left[
\Pi^{A_\ell}-\Pi^{\la A_{\ell-1}} \vb^{a_\ell\ra}
-\left({\ell+1\over 2\ell+3}\right)
\Pi^{A_\ell a}v_{\!_Ba}\right]
+{\cal O}[3]\,.
\label{nl9}
\end{eqnarray}
For $\ell=3$, the second term in square brackets on the right
of Eq. (\ref{nl9}) must be multiplied by ${3\over2}$.
The temperature fluctuation multipoles $\tau_{A_\ell}$ are determined
in principle from the radiation dynamic multipoles $\Pi_{A_\ell}$ via
equations (\ref{r29}) and (\ref{r30}).

{\em These equations show in a transparent
and explicitly 1+3 covariant and gauge-invariant form
precisely which physical effects are directly responsible for
the evolution of CMB anisotropies in an inhomogeneous universe.}
They show how the matter content of the universe generates
anisotropies. This happens directly through direct interaction
of matter with the radiation, as encoded
in the Thomson scattering terms on the right of equations
(\ref{nl5}), (\ref{nl6}), (\ref{nl8}) and (\ref{nl9}).
And it happens indirectly, as matter
generates inhomogeneities in the gravitational field via the
field equations (\ref{e1})--(\ref{c5})
and the evolution equation (\ref{t22}) for the baryonic velocity
$\vb^a$. This in turn feeds back into the multipole equations via the
kinematic quantities, the baryonic velocity $\vb^a$, and the
spatial gradient $\D_a\rho_{\!_R}$ in the dipole equation
(\ref{nl6}). The coupling of the multipole equations themselves
provides
an up and down cascade of effects, shown in general
by equation (\ref{nl9}).
Power is transmitted to the $\ell$-multipole by lower
multipoles through the dominant (linear) distortion term
$\D^{\la a_{\ell}} \Pi^{A_{\ell-1} \ra}$, as well as through
nonlinear terms coupled to the 4-acceleration
($A^{\la a_\ell}\Pi^{A_{\ell-1}\ra}$), baryonic velocity
($\vb^{\la a_\ell}\Pi^{A_{\ell-1}\ra}$), and shear
($\sigma^{\la a_\ell a_{\ell-1}}\Pi^{A_{\ell-2}\ra}$).
Simultaneously, power cascades down from higher multipoles
through the linear divergence term $(\div\Pi)^{A_\ell}$,
and the nonlinear terms coupled to $A^a$, $\vb^a$ and
$\sigma^{ab}$. (Note that the vorticity coupling does not
transmit across multipole levels.)

The equations for the radiation (and neutrino) multipoles
generalize the equations given by Challinor and
Lasenby \cite{cl2}, to which they
reduce when we remove all terms $O(\epsilon v_{\!_B})$
and $O(\epsilon^2)$. In this case, i.e. FLRW-linearization,
there is major simplification of the equations:
\begin{eqnarray}
\dot{\rho}_{\!_R}+{\ts{4\over3}}\Theta\rho_{\!_R}+\div q_{\!_R}
&\approx& 0\,,
\label{nl5a}\\
\dot{q}_{\!_R}^{ a}+4H q_{\!_R}^a
+{\ts{4\over3}}\rho_{\!_R}A^a+{\ts{1\over3}}\D^a\rho_{\!_R}
+(\div\pi_{\!_R})^a
&\approx&\ne\st\left({\ts{4\over3}}\rho_{\!_R}\vb^a-q_{\!_R}^a
\right)\,,
\label{nl6a}\\
\dot{\pi}_{\!_R}^{ ab}+4H \pi_{\!_R}^{ab}
+{\ts{8\over15}}\rho_{\!_R}\sigma^{ab}+
{\ts{2\over5}}\D^{\la a}q_{\!_R}^{b\ra}
+{8\pi\over35}\left(\div \Pi\right)^{ab}
&\approx&
-{\ts{9\over10}}\ne\st\pi_{\!_R}^{ab}\,,
\label{nl8a}
\end{eqnarray}
and for $\ell\geq3$
\be
\dot{\Pi}^{A_\ell}+4H \Pi^{A_\ell}+
\D^{\la a_\ell}\Pi^{A_{\ell-1}\ra}
+{(\ell+1)\over(2\ell+3)}\left(\div \Pi\right)^{A_\ell}
\approx
-\ne\st\Pi^{A_\ell}\,.
\label{nl9a}
\ee
These linearized equations, together with the linearized
equations governing the kinematic and free gravitational
quantities, given by equations (\ref{e1})--(\ref{c5}) with zero right
hand sides, may be covariantly split into scalar, vector and
tensor modes, as described in \cite{bde,cl,cl2}.
The modes can then be expanded in covariant eigentensors
of the comoving Laplacian, and the Fourier coefficients obey
ordinary differential equations, facilitating
numerical integration. Such integrations are performed for
scalar modes by Challinor and Lasenby \cite{cl2},
with further analytical results given in \cite{cl,cl2,ge2,c}.

However, in the nonlinear case, it is no longer possible
to split into scalar, vector and tensor modes \cite{pc,mm2,bmms}.
A simple illustration of this arises in dust spacetimes, which may
be considered as a simplified model after last scattering
if we neglect the dynamical effects of baryons, radiation and
neutrinos.
If one attempts to carry over
the linearized scalar-mode conditions \cite{bde,cl2}
\[
\omega_a=0=H_{ab}\,,
\]
into the nonlinear regime, it turns out that a non-terminating
chain of integrability conditions must be satisfied, so that
the models are in general inconsistent unless they have high
symmetry \cite{eulem,mle}. Thus, even in this simple case, it
is not possible to isolate scalar modes. In particular,
gravitational radiation, with $\c H_{ab}\neq0$ (see
\cite{dbe2,he,mes2}), must in general be present.

The generalized equations given above can form the basis for
investigating
the implications of nonlinear dynamical effects in general, and
second-order effects against an FLRW background in particular. More
quantitative and detailed investigations along these lines are
taken up in further papers. Here we will confine
ourselves to a qualitative analysis.

\subsection{Nonlinear effects on kinematic, gravitational
and dynamic quantities}

Evolution of the expansion of the universe $\Theta$, given by equation
(\ref{e2}), is retarded by the nonlinear shear term
$-\sigma_{ab}\sigma^{ab}$,
and accelerated by the nonlinear vector terms $+A_aA^a$ and
$+2\omega_a\omega^a$ (see also \cite{ell}).
The vorticity evolution equation (\ref{e4}) has a nonlinear
coupling $\sigma_{ab}\omega^b$
of vorticity to shear, whose effect will depend
on the alignment of vorticity relative to the
shear eigendirections. The shear evolution equation (\ref{e5})
has tensor-tensor and vector-vector type
couplings, which are the tensor counterpart of similar
terms in the expansion evolution. But in addition, relative
velocity effects enter via the total anisotropic stress term.
From equation (\ref{nl4}), we see that baryonic and cold dark matter
contributions of the form $\rho v^{\la a}v^{b\ra}$
to the shear evolution arise at the nonlinear level.
The constraint equations (\ref{c1}) and (\ref{c2}) show that
acceleration and vorticity provide scalar ($A^a\omega_a$)
and vector ($[\omega,A]_a$) nonlinear source terms for
respectively the vorticity and shear.

The free gravitational 
fields, which 1+3 covariantly describe tidal forces and
gravitational radiation (see \cite{ell,haw,dbe2,he,mes2}),
and therefore in particular
control the tensor contribution to CMB anisotropies,
are governed by the Maxwell-like equations (\ref{e6}), (\ref{e7}),
(\ref{c4}) and (\ref{c5}). This is the foundation for the
electromagnetic analogy. The role of nonlinear coupling terms
in these equations is more complicated -- see \cite{mb2} for
a full discussion. Here
we note that nonlinear couplings of the shear and vorticity
to the energy flux and gravito-magnetic field act as source terms
for the gravito-electric field -- see Eq. (\ref{c4}),
while nonlinear couplings of the shear and vorticity
to the anisotropic stress and gravito-electric field act as
source terms for the gravito-magnetic field -- see Eq. (\ref{c5}).

From equations (\ref{t19})--(\ref{t22}), we see that
for baryonic and cold dark
matter, nonlinear relative velocity terms act as
a source for the linear parts of the
evolution equations for energy density and relative velocity.
While the 4-acceleration $A_a$ is involved in
correction terms in all these equations, the vorticity $\omega_a$ and
shear $\sigma_{ab}$ only enter nonlinear corrections
of the velocity equations, and not the energy density equations.
This reflects the fact that
vorticity and shear are volume-preserving.
The kinematic corrections to the evolution
of matter relative velocity are of the form
$A_av^a$, $[\omega,v]_a$ and $\sigma_{ab}v^b$.
For
the massless species, as shown by equations (\ref{r26c}), (\ref{r26d})
and (\ref{nl5}), (\ref{nl6}),
the same form of corrections arises in the energy flux evolution,
since energy flux is of the form ${4\over3}\rho v^a$ when
the photon and neutrino frames are chosen as the energy frame.
Vorticity also does not affect
energy density, but shear does, owing to the
intrinsic anisotropic stress
of photons and neutrinos, which couples with the shear.

Baryonic and radiation conservation equations are both affected
by nonlinear Thomson correction terms, which involve
a coupling of the baryonic relative velocity $\vb^a$ to
the radiation energy density, momentum density and anisotropic
stress. In particular, we note that there is a
{\em nonzero energy density transfer}
due to Thomson scattering at second-order.

\subsection{Nonlinear effects on radiation multipoles}

Nonlinear Thomson scattering corrections also affect the evolution
of the radiation quadrupole $\pi_{\!_R}^{ab}$, as shown by
equation (\ref{nl8}). In this case, the baryonic relative
velocity couples to the radiation dipole $q_{\!_R}^a$ and
octopole $\Pi^{abc}$. Note also the ${1\over 10}$
correction to the linear Thomson term $\ne\st\pi_{\!_R}^{ab}$,
in agreement with \cite{HS95a,cl2}.
This correction arises from incorporating anisotropic effects in the
scattering integral (while neglecting polarization effects,
as noted earlier).

The general evolution equation (\ref{nl9}) for the
radiation dynamic multipoles $\Pi^{A_\ell}$ shows that
{\em five successive multipoles,}
i.e. for $\ell-2$, $\cdots$, $\ell+2$, are linked together
in the nonlinear case. Furthermore, the 4-acceleration $A_a$
couples to the $\ell\pm 1$ multipoles, the vorticity $\omega_a$
couples to the $\ell$ multipole, and the shear $\sigma_{ab}$
couples to the $\ell\pm2$ and $\ell$ multipoles. All of these
couplings are nonlinear, except for $\ell=1$ in the case of
$A_a$, and $\ell=2$ in the case of $\sigma_{ab}$. These latter
couplings that survive linearization are shown in the dipole
equation (\ref{nl6}) (i.e. $\rho_{\!_R}A^a$)
and the quadrupole equation (\ref{nl8})
(i.e. $\rho_{\!_R}\sigma^{ab}$).
The latter term drives Silk damping during the
decoupling process \cite{mt2}.
Nonlinear corrections introduce additional acceleration and shear
terms. {\em Vorticity corrections are purely nonlinear,}
i.e. vorticity
has no direct effect at the linear level, and a linear approach could
produce the false impression that vorticity has no direct
effect at all
on the evolution of CMB anisotropies.
However, for very high $\ell$, i.e. on very small angular scales,
the nonlinear vorticity term could in principle be non-negligible.

The disappearance of most
of the kinematic terms upon linearization is further reflected in the
fact that the linearized equations link only {\em three}
successive moments, i.e. $\ell$, $\ell\pm1$. This is clearly
seen in equation (\ref{nl9a}).

In addition to $A_a$ and $\omega_a$, there is a further vector
coupling at the nonlinear level, i.e. the coupling of the baryonic
velocity $\vb^a$ to the $\ell\pm1$ multipoles in the Thomson
scattering source term of the evolution equation (\ref{nl9}).
In fact these
nonlinear velocity corrections are of precisely the same
tensorial form
as the acceleration corrections on the left hand side, only
with different weighting factors.
Linearization, by removing these terms, also has the
effect of removing the nonlinear contribution
of the radiation multipoles $\Pi^{A_{\ell\pm1}}$ to
the collision multipole $K^{A_\ell}$.

One notable feature of the nonlinear terms is that
some of them scale like $\ell$ for large $\ell$, as
already noted in the case of vorticity. There are
no purely linear terms with this property, which has
an important consequence, i.e.
that {\em for very high $\ell$ multipoles
(corresponding to very small angular scales in CMB observations),
certain nonlinear terms can reach the same order of magnitude
as the linear contributions.}
(Note that the same effect applies to the neutrino background.)
The relevant nonlinear terms in Eq. (\ref{nl9})
are (for $\ell\gg 1$):
\[
-\ell \left(
{\ts{1 \over 4}} \sigma_{bc} \Pi^{bc A_{\ell} }
+ \sigma^{\la a_{\ell} a_{\ell-1}} \Pi^{A_{\ell-2}\ra}
-A^{\la a_\ell}\Pi^{A_{\ell-1}\ra}+{\ts{1\over2}}A_b\Pi^{bA_\ell}
+\omega^b\ep_{bc}{}{}^
{\la a_\ell}\Pi^{A_{\ell-1}\ra c}\right)\,.
\]
The observable imprint of this effect will be made after last
scattering. In the free-streaming era, it is reasonable to
neglect the vorticity relative to the shear. We can remove the
acceleration term by choosing $u^a$ as the
dynamically dominant cold dark matter frame (i.e.
choosing $v_{\!_C}^a=0$), as in \cite{cl2}.
It follows from equations (\ref{r31}) and (\ref{nl9}) that the
nonlinear correction to the rate of change of the
linearized temperature fluctuation multipoles is
\be
\delta(\dot{\tau}^{A_\ell})\sim
\ell \left(
{\ts{1 \over 4}} \sigma_{bc} \tau^{bc A_{\ell} }
+ \sigma^{\la a_{\ell} a_{\ell-1}} \tau^{A_{\ell-2}\ra}\right)
~\mbox{ for }~\ell\gg1\,.
\label{r40}\ee
The linear solutions for $\tau_{A_\ell}$ and $\sigma_{ab}$
can be used
on the right hand side to estimate the
correction to second order.
Its effect on observed anisotropies will be estimated
by integrating $\delta(\dot{\tau}^{A_\ell})$
from last scattering to now.
(See \cite{GE} for the relation between
the $\tau_{A_\ell}$ and the angular correlations).
In the case of scalar perturbations, these solutions are
given by Challinor and Lasenby \cite{cl2} (see also
\cite{ge2}).

Finally, we note that the well-known Vishniac and
Rees-Sciama second-order effects also become significant
at high $\ell$, and can eventually dominate the linear
contributions to CMB anisotropies on small enough angular
scales (typically $\ell>10^3$ or more) \cite{HS95a}.

\subsection{Temperature fluctuation multipoles}
We can normalize the radiation dynamic multipoles $\Pi^{A_\ell}$
to define the dimensionless multipoles  ($\ell\geq1$)
\[
{\cal T}^{A_\ell}=\left({\pi\over rT^4}\right)\Pi^{A_\ell}
\approx\tau^{A_\ell}\,.
\]
Thus the ${\cal T}^{A_\ell}$ are equal to
the temperature fluctuation multipoles plus nonlinear corrections.
In terms of these quantities, the hierarchy of radiation
multipoles becomes:
\begin{eqnarray}
{\dot{T}\over T} &=& -{\ts{1\over3}}\Theta-{\ts{1\over3}}
\D_a{\cal T}^a
\nonumber\\
&&{}-{\ts{4\over3}}{\cal T}^a{\D_a T\over T}-{\ts{2\over3}}A_a
{\cal T}^a-{\ts{2\over15}}\sigma_{ab}{\cal T}^{ab}
+{\ts{1\over3}}\ne\st v_{\!_Ba}\left(\vb^a-{\cal T}^a\right)
+{\cal O}[3]\,,
\label{nl10}\\
\dot{{\cal T}}^a &=&-4\left({\dot T\over T}+{\ts{1\over3}}\Theta
\right){\cal T}^a
-{\D^aT\over T}-A^a
-{\ts{2\over5}}\D_b{\cal T}^{ab}
+\ne\st\left(\vb^a-{\cal T}^a\right)
\nonumber\\
&&{}+{\ts{2\over5}}\ne\st {\cal T}^{ab}v_{\!_Bb}
-\sigma^a{}_b{\cal T}^b
-{\ts{2\over5}}A_b{\cal T}^{ab}
+[\omega,{\cal T}]^a
-{\ts{8\over5}}{\cal T}^{ab}{\D_b T\over T}
+{\cal O}[3]\,,
\label{nl11}\\
\dot{{\cal T}}^{ab}  &=&
- 4 \left( {\dot{T}\over T} + {\ts{1 \over 3}}\Theta \right)
{\cal T}^{ab}
- \sigma^{ab} - \D^{\la a} {\cal T}^{b \ra}
- {\ts{3 \over 7}}\D_c {\cal T}^{abc} -{\ts{9\over10}}\ne\st
{\cal T}^{ab}
\nonumber \\
&&{}+\ne\st\left({\ts{1\over2}}{\cal T}^{\la a}\vb^{b\ra}+
{\ts{3\over7}}{\cal T}^{abc}v_{\!_Bc}
+{\ts{3\over4}}\vb^{\la a}\vb^{b\ra}
\right)-5A^{\la a}{\cal T}^{b\ra}
 - {\ts{4 \over 21}} \sigma_{cd}{\cal T}^{abcd}
 \nonumber\\
&&{}+ 2 \omega^c\ep_{cd}{}{}^{\la a}{\cal T}^{b\ra d}
- {\ts{10 \over 7} }\sigma_c{}^{\la a}{\cal T}^{b\ra c}
-{\ts{12\over7}}{\cal T}^{abc} {\D_c T\over T}
+{\cal O}[3]\,,
 \label{r38}
\end{eqnarray}
and, for $\ell>3$:
\begin{eqnarray}
\dot{\cal T}^{A_{\ell}} &=&-
 4 \left( { \dot{T}\over T} + {\ts{1 \over 3}}\Theta \right)
{\cal T}^{A_\ell}
-\D^{\la a_\ell} {\cal T}^{A_{\ell-1} \ra} -
{(\ell+1) \over (2 \ell+3)}\D_b {\cal T}^{b A_{\ell}}
-\ne\st{\cal T}^{A_\ell}
\nonumber\\
&&{}+\ne\st\left[{\cal T}^{\la A_{\ell-1}}\vb^{a_\ell\ra}+
\left({\ell+1\over 2\ell+3}\right){\cal T}^{A_\ell b}v_{\!_Bb}\right]
+{(\ell+1)(\ell-2) \over (2 \ell +3)}  A_b {\cal T}^{b A_{\ell}}
\nonumber\\
&&{}-(\ell +3) A^{\la a_\ell} {\cal T}^{A_{\ell-1} \ra}
+ \ell \omega^b \ep_{bc}{}{}^{( a_{\ell}}
{\cal T}^{A_{\ell-1}) c }
+(\ell+2) \sigma^{\la a_{\ell} a_{\ell-1}} {\cal T}^{A_{\ell-2} \ra}
\nonumber \\
&&{}+
{(\ell-1)(\ell+1)(\ell+2) \over (2\ell+3)(2\ell+5)}
\sigma_{bc} {\cal T}^{bc A_\ell}
- { 5 \ell \over (2 \ell +3)} \sigma_b{}^{\la a_{\ell}}
{\cal T}^{A_{\ell-1} \ra b}
 -4 {(\ell+1) \over (2 \ell +3)}{\cal T}^{A_\ell b}
 {\D_b T\over T} +{\cal O}[3]\,.
\label{r36}
\end{eqnarray}
For $\ell=3$, the
Thomson term ${\cal T}^{\la A_{\ell-1}}\vb^{a_\ell\ra}$ must
be multiplied by ${3\over2}$.

The nonlinear multipole equations given in this
form show more clearly the evolution of temperature
anisotropies (including the monopole, i.e. the average temperature
$T$). Althought the ${\cal T}^{A_\ell}$ only determine
the actual temperature
fluctuations $\tau^{A_\ell}$ to linear order, they are
a useful dimensionless measure of anisotropy. Furthermore,
equations (\ref{nl10})--(\ref{r36}) apply as the evolution
equations
for temperature  fluctuation multipoles when the
radiation anisotropy
is small (i.e.
${\cal T}^{A_\ell}=
\tau^{A_\ell}$), but the
spacetime inhomogeneity and anisotropy are not restricted.
This includes the particular case of small CMB anisotropies in
general Bianchi universes, or in perturbed Bianchi universes.

FLRW-linearization, i.e. the case when only first order effects
relative to the FLRW limit are considered, reduces the above
equations to:
\begin{eqnarray}
{\dot{T}\over T} &\approx& -{\ts{1\over3}}\Theta-{\ts{1\over3}}
\D_a{\tau}^a \,,
\label{nl10a}\\
\dot{{\tau}}^a &\approx&
-{\D^aT\over T}-A^a
-{\ts{2\over5}}\D_b{\tau}^{ab}
+\ne\st\left(\vb^a-{\tau}^a\right) \,,
\label{nl11a}\\
\dot{{\tau}}^{ab}  &\approx&
- \sigma^{ab} - \D^{\la a} {\tau}^{b \ra}
- {\ts{3 \over 7}}\D_c {\tau}^{abc} -{\ts{9\over10}}\ne\st
{\tau}^{ab}\,,
 \label{r38a}
\end{eqnarray}
and, for $\ell\geq3$:
\begin{eqnarray}
\dot{\tau}^{A_{\ell}} &\approx&
-\D^{\la a_\ell} {\tau}^{A_{\ell-1} \ra} -
{(\ell+1) \over (2 \ell+3)}\D_b {\tau}^{b A_{\ell}}
-\ne\st{\tau}^{A_\ell}\,.
\label{r36a}
\end{eqnarray}
These are the 1+3 covariant and gauge-invariant multipole
generalizations of the
Fourier mode formulation of the integrated Boltzmann equations used
in the standard literature (see e.g. \cite{HS95a} and the references
therein).
Equations (\ref{nl10a})--(\ref{r38a}) were given in \cite{mes}
in the free-streaming case $\ne=0$.

As noted before, there is still
a gauge freedom here associated with the choice of 4-velocity $u^a$.
Given any physical choice for this 4-velocity which tends to
the preferred 4-velocity in the FLRW limit,
the $\ell \ge 1$ equations are
gauge-invariant.

\section{Conclusions}

We have used a covariant Lagrangian approach, in which all
the relevant physical and geometric quantities occur directly
and transparently, as PSTF tensors measured in the comoving rest
space. There is no restriction
on the deviation of geometric and physical quantities from
FLRW limiting values, so that arbitrary nonlinear behavior may
in principle be treated.
We have derived the corresponding equations governing the
generation and evolution of inhomogeneities and CMB anisotropies
in nonlinear generality,
without a priori restrictions on spacetime geometry or
specific assumptions about early-universe particle physics,
structure formation history, etc. Thus we have developed a useful
approach to the analysis of local nonlinear effects in
CMB anisotropies,
with the clarity and transparency arising from 3+ 1 covariance.
The equations are readily linearized in a gauge-invariant way,
and then the methods of \cite{GE}
may be used to expand in scalar modes and
regain well-known first-order results \cite{ge2}
(see also \cite{cl,cl2,c}).

This approach allowed us to identify and qualitatively describe
some of the key local nonlinear effects, and more
quantitative results will be considered in further
papers.
We calculated the nonlinear form of Thomson scattering
multipoles (given the initial
simplifying assumption of no polarization), revealing the new
effect of coupling between the baryonic bulk velocity and
radiation brightness multipoles of order $\ell\pm1$. We also found
the nonlinear effects of relative velocities of particle
species on the dynamic quantities that source the gravitational
field. These effects also operate on the conservation equations,
including evolution equations for the relative velocities of
baryonic and cold dark matter.

Nonlinear effects come together in the hierarchy of evolution
equations for the radiation dynamic (brightness) multipoles,
which determine the CMB temperature anisotropies. In addition
to the nonlinear Thomson contribution, we identified nonlinear
couplings of the kinematic quantities to the multipoles of
order $\ell\pm2$, $\ell\pm 1$, and $\ell$. These quantities
themselves are governed by nonlinear evolution equations,
which provides part of the link between CMB anisotropies and
inhomogeneities in
the gravitational field and sources. The link is also carried
by the spatial gradient of radiation energy density
(equivalently,
average radiation all-sky temperature), and the baryonic relative
velocity. Furthermore, there is internal up- and down-transmission
of power within the multipole hierarchy, supported by the
kinematic couplings as well as by distortion and divergence
derivatives of the multipoles.

We used our analysis of the radiation multipoles to identify
new effects that operate at high $\ell$. In particular,
we showed that there is
a nonlinear shear correction effect on small angular
scales, whose impact on the angular power spectrum was
qualitatively described.
The quantitative analysis of this and other nonlinear effects
is a subject of further research.

\[ \]
{\bf Acknowledgements}

Special thanks to Anthony Challinor for very helpful comments
and criticisms. Thanks also to
Peter Dunsby, Bill Stoeger, Bruce Bassett, Henk van Elst,
Malcolm MacCallum, Arthur Kosowsky and David Matravers for
useful comments and discussions. This work was
supported by the South African Foundation for
Research and Development.

\appendix
\section{Exact nonlinear relative velocity equations}

Change in 4-velocity:
\be
\tilde{u}_a = \gamma(u_a+v_a) ~\mbox{ where }~
\gamma=(1-v^2)^{-1/2}\,,~v_au^a=0\,. \label{B25}
\ee
Change in fundamental algebraic tensors:
\begin{eqnarray}
\t{h}_{ab} &=& h_{ab}+\gamma^2\left[v^2u_a u_b+2u_{(a}v_{b)}
+v_a v_b\right] \,,\label{ab1}\\
\tilde{\ep}_{abc} &=& \gamma\ep_{abc}+\gamma\left\{2u_{[a}\ep_{b]cd}
+u_c\ep_{abd}\right\}v^d\,. \label{ab2}
\end{eqnarray}

Transformed kinematic quantities are defined by
\[
\nabla_b\t{u}_a={\ts{1\over3}}\t{\Theta}\t{h}_{ab}+\t{\sigma}_{ab}
+\t{\ep}_{abc}\t{\omega}^c-\t{A}_a\t{u}_b\,,
\]
which implies, using
$\nabla_a\gamma=\gamma^3v^b\nabla_av_b$,
and Eq. (\ref{a2}), the following kinematic transformations
\cite{maa2}:
\begin{eqnarray}
\t{\Theta} &=& \gamma\Theta+\gamma\left(\div v+A^av_a\right)
+\gamma^3W \,,\label{ab3}\\
\t{A}_a &=& \gamma^2A_a+\gamma^2\left\{
\dot{v}_{\la a\ra}+
{\ts{1\over3}}\Theta v_a+
\sigma_{ab}v^b-[\omega,v]_a
+\left({\ts{1\over3}}\Theta v^2+A^bv_b+\sigma_{bc}v^bv^c
\right)u_a
\right.\nonumber\\
&&\left.
+{\ts{1\over3}}(\div v)v_a+{\ts{1\over2}}[v,\c v]_a
+v^b\D_{\la b}v_{a\ra}\right\}
+\gamma^4W(u_a+v_a) \,, \label{ab4}
\\
\t{\omega}_a &=& \gamma^2\left\{\left(1-{\ts{1\over2}}v^2\right)
\omega_a- {\ts{1\over2}}\c v_a+{\ts{1\over2}}v_b\left(2\omega^b-\c v^b
\right)u_a+{\ts{1\over2}}v_b\omega^bv_a \right.
\nonumber\\
&&\left.{}+{\ts{1\over2}}[A,v]_a+{\ts{1\over2}}[\dot{v},v]_a
+{\ts{1\over2}}\ep_{abc}\sigma^b{}_dv^cv^d \right\} \,, \label{ab5}\\
\t{\sigma}_{ab} &=& \gamma\sigma_{ab}
+\gamma(1+\gamma^2)u_{(a}\sigma_{b)c}v^c
+\gamma^2A_{(a}\left[v_{b)}+v^2u_{b)}\right] \nonumber\\
&&{}+ \gamma\D_{\la a}v_{b\ra}
-{\ts{1\over3}}h_{ab}\left[
A_cv^c+\gamma^2\left(W-\dot{v}_cv^c\right)\right]
\nonumber\\
&&{}+\gamma^3u_au_b\left[\sigma_{cd}v^cv^d+{\ts{2\over3}}v^2A_cv^c
-v^cv^d\D_{\la c}v_{d\ra}+\left(\gamma^4-{\ts{1\over3}}v^2\gamma^2
-1\right)W\right]\nonumber\\
&&{}+\gamma^3u_{(a}v_{b)}\left[A_cv^c+\sigma_{cd}v^cv^d-\dot{v}_cv^c+
2\gamma^2\left(\gamma^2-{\ts{1\over3}}\right)W\right] \nonumber\\
&&{}+{\ts{1\over3}}\gamma^3v_av_b\left[\div v-A_cv^c
+\gamma^2\left(3\gamma^2-1\right)W\right]+\gamma^3v_{\la a}
\dot{v}_{b\ra}+v^2\gamma^3u_{(a}\dot{v}_{\la b\ra)} \nonumber\\
&&{}+\gamma^3v_{(a}\sigma_{b)c}v^c-\gamma^3[\omega,v]_{(a}
\left\{v_{b)}+v^2u_{b)}\right\}
+2\gamma^3v^c\D_{\la c}v_{(a\ra}\left\{v_{b)}+u_{b)}\right\}     \,,
\label{ab6}
\end{eqnarray}
where
\[
W\equiv
\dot{v}_cv^c+{\ts{1\over3}}v^2\div v+v^cv^d\D_{\la c}v_{d\ra} \,.
\]

Transformed dynamic quantities are \cite{maa2}:
\begin{eqnarray}
\t{\rho} &=& \rho+\gamma^2\left[v^2(\rho+p)-2q_av^a+\pi_{ab}v^av^b
\right] \,,\label{ab7}\\
\t{p} &=& p+{\ts{1\over3}}\gamma^2\left[v^2(\rho+p)-2q_av^a
+\pi_{ab}v^av^b\right] \,,\label{ab8}\\
\t{q}_a &=& \gamma q_a-\gamma\pi_{ab}v^b-\gamma^3\left[(\rho+p)
-2q_bv^b+\pi_{bc}v^bv^c\right]v_a \nonumber\\
{}&&-\gamma^3\left[v^2(\rho+p)-(1+v^2)q_bv^b+\pi_{bc}v^bv^c\right]u_a
\,, \label{ab9} \\
\t{\pi}_{ab} &=& \pi_{ab}+2\gamma^2v^c\pi_{c(a}\left\{u_{b)}+
v_{b)}\right\}-2v^2\gamma^2q_{(a}u_{b)}-2\gamma^2q_{\la a}v_{b\ra}
\nonumber\\
{}&&-{\ts{1\over3}}\gamma^2\left[v^2(\rho+p)+\pi_{cd}v^cv^d
\right]h_{ab} \nonumber\\
&&{}+{\ts{1\over3}}\gamma^4\left[2v^4(\rho+p)-4v^2q_cv^c
+(3-v^2)\pi_{cd}v^cv^d\right]u_au_b \nonumber\\
{}&&+{\ts{2\over3}}\gamma^4\left[2v^2(\rho+p)-(1+3v^2)q_cv^c+2
\pi_{cd}v^cv^d\right]u_{(a}v_{b)} \nonumber\\
{}&&+{\ts{1\over3}}\gamma^4\left[(3-v^2)(\rho+p)-4q_cv^c+2
\pi_{cd}v^cv^d\right]v_av_b \,. \label{ab10}
\end{eqnarray}

gravito-electric/magnetic field: using \cite{maa}
\begin{eqnarray*}
C_{ab}{}{}^{cd} &=&
4\left\{u_{[a}u^{[c}+h_{[a}{}^{[c}\right\}E_{b]}{}^{d]}
+2\ep_{abe}u^{[c}H^{d]e}+2u_{[a}H_{b]e}\ep^{cde} \\
&=&
4\left\{\tilde{u}_{[a}\tilde{u}^{[c}+\tilde{h}_{[a}{}^{[c}\right\}
\tilde{E}_{b]}{}^{d]}
+2\tilde{\ep}_{abe}\tilde{u}^{[c}\tilde{H}^{d]e}+2
\tilde{u}_{[a}\tilde{H}_{b]e}\tilde{\ep}^{cde}  \,,
\end{eqnarray*}
we find the transformation \cite{maa2}:
\begin{eqnarray}
\tilde{E}_{ab} &=&
\gamma^2\left\{(1+v^2)
E_{ab}+v^c\left[2\ep_{cd(a}H_{b)}{}^d+
2E_{c(a}u_{b)}\right.\right.\nonumber\\
&&\left.\left.{}+(u_au_b+h_{ab})E_{cd}v^d-2E_{c(a}v_{b)}
+2u_{(a}\ep_{b)cd}H^{de}v_e\right]\right\} \,, \label{ab11}\\
\tilde{H}_{ab} &=&
\gamma^2\left\{(1+v^2)
H_{ab}+v^c\left[-2\ep_{cd(a}E_{b)}{}^d+
2H_{c(a}u_{b)}\right.\right.\nonumber\\
&&\left.\left.{}+(u_au_b+h_{ab})H_{cd}v^d-2H_{c(a}v_{b)}
-2u_{(a}\ep_{b)cd}E^{de}v_e\right]\right\}\,. \label{ab12}
\end{eqnarray}
This may be compared with the electromagnetic transformation
\begin{eqnarray*}
\tilde{E}_a &=& \gamma\left\{ E_a+[v,H]_a+v^bE_bu_a\right\}\,,\\
\tilde{H}_a &=& \gamma \left\{ H_a-[v,E]_a+v^bH_bu_a\right\}\,,
\end{eqnarray*}
where
\[
F_{ab} = 2u_{[a}E_{b]}+\ep_{abc}H^c
= 2\tilde{u}_{[a}\tilde{E}_{b]}
+\tilde{\ep}_{abc}\tilde{H}^c\,.
\]

Note that all the transformations above are
given explicitly in terms of irreducible
quantities (i.e. irreducible in the original $u^a$-frame).



\begin{references}

\bi{ehl}
J. Ehlers, Gen. Relativ. Gravit. {\bf 25}, 1225 (1993)
(translation of 1961 article).

\bi{ell}
G.F.R. Ellis, in {\em General Relativity and Cosmology}, edited by
R.K. Sachs (Academic, New York, 1971).

\bi{haw}
S.W. Hawking, Astrophys. J. {\bf 145}, 544 (1966).

\bibitem{eb}
G.F.R. Ellis and M. Bruni, Phys. Rev. D {\bf 40}, 1804 (1989).

\bibitem{ETMa}
G.F.R. Ellis, D.R. Matravers, and R. Treciokas, Ann.
Phys. (N.Y.) {\bf 150}, 455 (1983).

\bibitem{ETMb}
G.F.R. Ellis, R. Treciokas, and D.R. Matravers, Ann.
Phys. (N.Y.) {\bf 150}, 487 (1983).

\bibitem{EGS}
J. Ehlers, P. Geren, and R.K. Sachs,
J. Math. Phys. {\bf 9}, 1344 (1968).

\bibitem{TE}
R. Treciokas and G.F.R. Ellis,
Commun. Math. Phys. {\bf 23}, 1 (1971).

\bibitem{T2}
K.S. Thorne, Mon. Not. R. Astron. Soc. {\bf 194}, 439
(1981).

\bibitem{mes}
R. Maartens, G.F.R. Ellis, and W.R. Stoeger,
Phys. Rev. D {\bf 51}, 1525 (1995).

\bibitem{SW}
R.K. Sachs and A.M. Wolfe, Astrophys. J. {\bf 147}, 73 (1967).

\bi{rs}
M. Rees and D.W. Sciama, Nature (London) {\bf 519}, 611 (1968).

\bibitem{PY}
P.J.E. Peebles and J.T. Yu, Astrophys. J. {\bf 162}, 815 (1970).

\bi{sz2}
R.A. Sunyaev and Ya.B. Zeldovich, Astrophys. Space Sci. {\bf 7}, 3 (1970).

\bi{gz}
L.P. Grishchuk and Ya.B. Zeldovich, Astron. Zh. {\bf 55}, 209 (1978)
[Sov. Astron. {\bf 22}, 125 (1978)].

\bi{bar}
J. Bardeen, Phys. Rev. D {\bf 22}, 1882 (1980).

\bi{ks}
H. Kodama and M. Sasaki, Prog. Theor. Phys. {\bf 78}, 1 (1984).

\bi{lif}
E.M. Lifshitz, J. Phys. (Moscow) {\bf 10}, 116 (1946).

\bibitem{HS95a}
W. Hu and N. Sugiyama, Astrophys. J. {\bf 444}, 489 (1995).

\bibitem{HS95b}
W. Hu and N. Sugiyama, Phys. Rev. D {\bf 51}, 2599 (1995).

\bibitem{HW}
W. Hu and M. White, Astron. Astrophys. {\bf 315}, 33 (1996).

\bibitem{MB}
C.P. Ma and E. Bertschinger, Astrophys. J. {\bf 455}, 7 (1995).

\bibitem{sz}
S. Seljak and M. Zaldarriaga, Astrophys. J. {\bf 469}, 437 (1996).

\bibitem{zs}
M. Zaldarriaga and U. Seljak, Phys. Rev. D {\bf 55}, 1830 (1997).

\bi{zsb}
M. Zaldarriaga, U. Seljak, and E. Bertschinger, Astrophys. J.
{\bf 494}, 491 (1998).

\bi{hswz}
W. Hu, U. Seljak, M. White, and M. Zaldarriaga, Phys. Rev. D
{\bf 57}, 3290 (1998).

\bi{dk}
R. Durrer and T. Kahniashvili, Helv. Phys. Acta {\bf 71}, 445 (1998).

\bi{gs}
E. Gawiser and J. Silk, Science {\bf 280}, 1405 (1998).

\bi{hrlg}
S. Hancock, G. Rocha, A.N. Lasenby, and C.M. Gutierrez,
Mon. Not. R. Astron. Soc. {\bf 294}, L1 (1998).

\bi{sme}
W.R. Stoeger, R. Maartens, and G.F.R. Ellis,
Astrophys. J. {\bf 443}, 1 (1995).

\bibitem{mes3}
R. Maartens, G.F.R. Ellis, and W.R. Stoeger,
Phys. Rev. D {\bf 51}, 5942 (1995).

\bibitem{mes4}
R. Maartens, G.F.R. Ellis, and W.R. Stoeger,
Astron. Astrophys. {\bf 309}, L7 (1996).

\bibitem{SAG}
W.R. Stoeger, M. Araujo, and T. Gebbie,
Astrophys. J. {\bf 476}, 435 (1997).

\bibitem{dunsby}
P.K.S. Dunsby, Class. Quantum Grav. {\bf 14}, 3391 (1997).

\bibitem{cl}
A.D. Challinor and A.N. Lasenby, Astrophys. J., in press (1998).
(astro-ph/9804150)

\bi{cl2}
A.D. Challinor and A.N. Lasenby, Phys. Rev. D {\bf 58},
023001 (1998).

\bibitem{GE}
T. Gebbie and G.F.R. Ellis, astro-ph/9804316.

\bi{c}
A.D. Challinor, Ph.D. thesis, University of Cambridge (1998).

\bibitem{ehb}
G.F.R. Ellis, J. Hwang, and M. Bruni, Phys. Rev.
D {\bf 40}, 1819 (1989).

\bibitem{ebh}
G.F.R. Ellis, M. Bruni, and J. Hwang, Phys. Rev.
D {\bf 42}, 1035 (1990).

\bi{ls}
D.H. Lyth and E.D. Stewart, Astrophys. J. {\bf 361}, 343 (1990).

\bi{dun2}
P.K.S. Dunsby, Class. Quantum Grav. {\bf 8}, 1785 (1991).

\bibitem{bde}
M. Bruni, P.K.S. Dunsby, and G.F.R. Ellis,
Astrophys. J {\bf 395}, 34 (1992).

\bi{bed}
M. Bruni, G.F.R. Ellis, and P.K.S. Dunsby, Class. Quantum Grav.
{\bf 9}, 921 (1992).

\bibitem{dbe}
P.K.S. Dunsby, M. Bruni, and G.F.R. Ellis,
 Astrophys. J {\bf 395}, 54 (1992).

\bi{bggmv}
R. Brustein, M. Gasperini, M. Giovannini, V.F. Mukhanov, and G.
Veneziano, Phys. Rev. D {\bf 51}, 6744 (1995).

\bi{ber}
E. Bertschinger, in {\em Cosmology and Large-scale Structure},
edited by R. Schaeffer, J. Silk, M. Spiro, and V. Zinn-Justin
(Elsevier, Amsterdam, 1996)

\bi{mt}
R. Maartens and J. Triginer, Phys. Rev. D {\bf 56}, 4640 (1997).

\bi{tb}
C.G. Tsagas and J.D. Barrow, Class. Quantum Grav.
{\bf 14}, 2539 (1997); {\bf 15}, 3523 (1998).

\bi{mt2}
R. Maartens and J. Triginer, Phys. Rev. D {\bf 58}, 123507 (1998).


\bi{vee}
H. van Elst and G.F.R. Ellis, Class. Quantum
Grav. {\bf 15}, 3545 (1998).

\bi{maa2}
R. Maartens, Phys. Rev. D {\bf 58}, 124006 (1998). (astro-ph/9808235)

\bibitem{bm}
J.D. Barrow and R. Maartens, Phys. Rev. D {\bf 59}, 043502 (1999).

\bi{mtm}
R. Maartens, J. Triginer, and D.R. Matravers, astro-ph/9901213.

\bi{dbe2}
P.K.S. Dunsby, B.A. Bassett, and G.F.R. Ellis,
Class. Quantum Grav. {\bf 14}, 1215 (1997).

\bi{he}
P.A. Hogan and G.F.R. Ellis, Class. Quantum Grav.
{\bf 14}, A171 (1997).

\bibitem{mes2}
R. Maartens, G.F.R. Ellis, and S.T.C. Siklos,
Class. Quantum Grav. {\bf 14}, 1927 (1997).

\bibitem{mb2}
R. Maartens and B.A. Bassett, Class. Quantum Grav.
{\bf 15}, 705 (1998).

\bi{maa3}
B.A. Bassett, Phys. Rev. D {\bf 56}, 3439 (1997).

\bi{ed}
G.F.R. Ellis and P.K.S. Dunsby, in {\em Current Topics in
Astrofundamental Physics} edited by N. Sanchez (Kluwer, Dordrecht, 1998).

\bi{br}
A. Bonanno and V. Romano, Phys. Rev. D {\bf 49}, 6450 (1994).

\bi{emn}
G.F.R. Ellis, R. Maartens, and S.D. Nel, Mon. Not. R.
Astron. Soc. {\bf 184}, 439 (1978).

\bi{pc}
T. Pyne and S.M. Carroll, Phys. Rev. D {\bf 53}, 2920 (1996).

\bi{mm2}
S. Mollerach and S. Matarrese, Phys. Rev. D {\bf 56}, 4494 (1997).

\bi{W83}
M.L. Wilson, Astrophys. J. {\bf 273}, 2 (1983).

\bi{ge2}
T. Gebbie, P.K.S. Dunsby, and G.F.R. Ellis (in preparation).

\bi{maa}
R. Maartens, Phys. Rev. D {\bf 55}, 463 (1997).

\bi{kt}
W. Kundt and M. Tr\"umper, Akad. Wiss. Lit. Mainz Abh.
Math. Naturwiss. Kl. {\bf 12}, 1 (1962).

\bi{ve}
H. van Elst, Ph.D. thesis, University of London, 1996.

\bi{mac}
M.A.H. MacCallum, gr-qc/9806003.

\bibitem{Is}
W. Israel and J.M. Stewart, Ann. Phys. (N.Y.) {\bf 118}, 341 (1979).

\bi{pl}
D. Psaltis and F.K. Lamb, Astrophys. J. {\bf 488}, 881 (1998).

\bi{bmms}
M. Bruni, S. Matarrese, S. Mollerach, and S. Sonego, Class. Quantum
Grav. {\bf 14}, 2585 (1997).

\bibitem{LQ}
R.W. Lindquist, Ann. Phys. (N.Y.) {\bf 37}, 487 (1966).

\bi{Ehlers}
J. Ehlers, in {\it General Relativity and Cosmology}, edited by
 R.K. Sachs (Academic, New York, 1971).

\bibitem{Stewart}
J.M. Stewart, {\it Non-Equilibrium Relativistic
Kinetic Theory} (Springer-Verlag, Berlin, 1971).

\bibitem{dGvL}
S.L. de Groot, W.A. van Leeuwen, and C.G. van Weert,
{\it Relativistic Kinetic Theory} (North-Holland, Amsterdam, 1980).

\bibitem{Br}
J. Bernstein, {\it Kinetic Theory in the Expanding Universe}
(Cambridge University Press, Cambridge, England, 1988).

\bi{wey}
R. Weymann, Astrophys. J. {\bf 145}, 560 (1966).

\bi{uza}
J.-P. Uzan, Class. Quantum Grav. {\bf 15}, 1063 (1998).

\bi{sil}
J. Silk, Astrophys. J. {\bf 151}, 459 (1968).

\bi{wei}
S.W. Weinberg, Astrophys. J. {\bf 168}, 175 (1971).

\bi{hssw}
W. Hu, D. Scott, N. Sugiyama, and M. White, Phys. Rev. D
{\bf 52}, 5498 (1995).

\bi{gd}
T. Gebbie, P.K.S. Dunsby, and G.F.R. Ellis (in preparation).

\bi{eulem}
H. van Elst, C. Uggla, W.M. Lesame, G.F.R. Ellis, and
R. Maartens, Class. Quantum Grav. {\bf 14}, 1151 (1997).

\bi{mle}
R. Maartens, W.M. Lesame, and G.F.R. Ellis, Class. Quantum Grav.
{\bf 15}, 1005 (1998).

\end{references}
\end{document}